\documentclass[10pt,journal,draftcls,letterpaper,twoside,onecolumn]{IEEEtran}

\usepackage{dsfont}
\usepackage{amsmath}
\usepackage{amssymb}
\usepackage{amsfonts}
\usepackage{graphicx}
\usepackage{amsmath}
\usepackage{multirow}
\usepackage{subfigure}
\usepackage{comment}
\usepackage{algorithm}
\usepackage{algorithmic}
\usepackage{color} 

\def\bfe{{\mathbf{e}}}

\def\bfh{{\mathbf{h}}}

\def\bfu{{\mathbf{u}}}
\def\bfv{{\mathbf{v}}}

\def\bfx{{\mathbf{x}}}
\def\bfy{{\mathbf{y}}}
\def\bfz{{\mathbf{z}}}

\def\bfH{{\mathbf{H}}}
\def\bfI{{\mathbf{I}}}

\def\bfW{{\mathbf{W}}}
\def\bfX{{\mathbf{X}}}

\def\bbE{{\mathbb{E}}}

\def\bbP{{\mathbb{P}}}

\def\dsE{{\mathds{E}}}

\def\calU{{\mathcal{U}}}

\def\calS{{\mathcal{S}}}

\def\calB{{\mathcal{B}}}

\def\calG{{\mathcal{G}}}

\def\calI{{\mathcal{I}}}

\def\calN{{\mathcal{N}}}

\def\calS{{\mathcal{S}}}

\def\calU{{\mathcal{U}}}

\newcommand{\Vobs}{\mathbf{y}}
\newcommand{\obs}[1]{y_{#1}}

\newcommand{\MATima}{\bfX}
\newcommand{\Vima}{\bfx}
\newcommand{\ima}[1]{x_{#1}}

\newcommand{\dimm}{M}
\newcommand{\dimn}{P}
\newcommand{\dimima}{n}

\newcommand{\ftrans}[2]{T\left(#1,#2\right)}
\newcommand{\MATtrans}{\mathbf{H}}
\newcommand{\Vtrans}[1]{\mathbf{h}_{i}}

\newcommand{\psf}{\boldsymbol{\kappa}}
\newcommand{\Vpsfparam}{\boldsymbol{\lambda}}
\newcommand{\psfparam}[1]{\lambda_{#1}}

\newcommand{\Vnoise}{\mathbf{n}}
\newcommand{\noisevar}{{\sigma^2}}

\newcommand{\hypervect}{\boldsymbol{\Phi}}
\newcommand{\paramvect}{\boldsymbol{\theta}}

\newcommand{\Valpha}{\boldsymbol{\alpha}}

\newcommand{\sample}[2]{#1^{(#2)}}
\newcommand{\samplebis}[2]{{#1}^{(#2)}}
\newcommand{\samplenoisevar}[1]{{\widetilde{\sigma}}^{2(#1)}}

\newcommand{\norm}[1]{\left\|#1\right\|}

\newcommand{\R}{\mathds{R}}

\newcommand{\dirac}[1]{\delta\left({#1}\right)}

\newcommand{\transp}{^T}

\newcommand{\Vzero}{\boldsymbol{0}}
\newcommand{\Id}[1]{\textbf{I}_{#1}}
\newcommand{\Indicfun}[2]{\textbf{1}_{#1}\left(#2\right)}

\newcounter{algo}
\renewcommand{\thealgo}{\arabic{algo}}

\newcommand{\figwidth}{0.7\columnwidth}
\newcommand{\figwidthhalf}{0.45\columnwidth}

\newcommand{\figwidthbis}{0.6\columnwidth}
\newcommand{\figwidthsmall}{0.23\textwidth}
\newcommand{\figwidthbig}{0.23\textwidth}

\newcommand{\figextension}{png}
\graphicspath{{figures/}}
\title{Semi-blind Sparse Image Reconstruction \\with Application to MRFM}

\author{Se Un Park, Nicolas Dobigeon and Alfred O. Hero
\thanks{Se Un Park and Alfred O. Hero are with University of Michigan, Department of EECS, Ann Arbor, MI
48109-2122, USA.
 (e-mail: \{seunpark,hero\}@umich.edu).}
\thanks{Nicolas Dobigeon is with University of Toulouse, IRIT/INP-ENSEEIHT, 2 rue
Camichel, BP 7122, 31071 Toulouse cedex 7, France. (e-mail: nicolas.dobigeon@enseeiht.fr).}
\thanks{This research was partially supported by a grant from ARO, grant number W911NF-05-1-0403.}
}

\begin{document}

\maketitle

\hyphenation{hie-rar-chi-cal as-tro-no-mi-cal}

{
\begin{abstract}
We propose a solution to the image deconvolution problem where the convolution kernel or point spread function (PSF) is
assumed to be only partially known. Small perturbations generated from the model are exploited to produce a few
principal components explaining the PSF uncertainty in a high dimensional space. Unlike recent developments on blind
deconvolution of natural images, we assume the image is sparse in the pixel basis, a natural sparsity arising in
magnetic resonance force microscopy (MRFM). Our approach adopts a Bayesian Metropolis-within-Gibbs sampling framework.
The performance of our Bayesian semi-blind algorithm for sparse images is
 superior to previously proposed semi-blind algorithms such as the
alternating minimization (AM) algorithm and blind algorithms developed for natural images.
 We illustrate our myopic algorithm on real MRFM tobacco virus data.
\end{abstract}
}

\begin{keywords}
Semi-blind (myopic) sparse deconvolution, Bayesian inference, Markov Chain Monte Carlo (MCMC) methods, MRFM experiment
\end{keywords}

\section{Introduction}
\label{sec:intro}

Recently, a new 3D imaging technology called magnetic resonance force microscopy (MRFM) has been developed. The
principles of MRFM were introduced by Sidles \cite{Sidles1991, Sidles1992, Sidles1995} who described its potential for
achieving 3D atomic scale resolution. In 1992 and 1996, Rugar \emph{et al.} \cite{Rugar1992,Zuger1996} reported
experiments that demonstrated the feasibility of MRFM and produced the first MRFM images. More recently, MRFM
volumetric spatial resolutions of less than $10nm$ have been demonstrated for imaging a biological sample
\cite{Degen2009}. The signal provided by MRFM is a so-called force map that is the $3$D convolution of the atomic spin
distribution and the point spread function (PSF) \cite{Chao2004}. This formulation casts the estimation of the spin
density from the force map as an inverse problem. Several approaches have been proposed to solve this inverse problem,
i.e., to reconstruct the unknown image from the measured force map. Basic algorithms rely on Wiener filters
\cite{Zuger1993,Zuger1994,Zuger1996} whereas others are based on iterative least squares reconstruction approaches
\cite{Chao2004,Degen2009,Degen2009compl}. More recently, this problem has been addressed within the Bayesian estimation
framework \cite{Ting2009,Dobigeon2009a}.

However, all of these reconstruction techniques require prior knowledge of the device response, namely the PSF. As
shown by Mamin {\em et al.} \cite{Mamin2003}, this PSF is a function of several parameters specified by the physics of
the device including: mass of cantilever probe, ferromagnetic constant of probe tip, and external field strength.
Unfortunately, in practice the physical parameters that tune the response of the MRFM tip are only partially known. In
such circumstances, the PSF used in the reconstruction algorithm is mismatched to the true PSF of the microscope and
the quality of standard MRFM image reconstruction will suffer if one does not account for this mismatch. Estimating the
unknown image and the partially known PSF jointly is usually referred to as semi-blind \cite{Makni2004, Pillonetto2007}
or myopic \cite{Sarri1998, Chenegros2007} deconvolution, and this is the approach taken in this paper. The myopic image
restoration problem was previously studied within a hierarchical Bayesian framework \cite{Molina1994} with partially
known blur functions in many applications, including natural and astronomical imaging
\cite{Galatsanos2000,Galatsanos2002}. This previous work \cite{Galatsanos2000,Galatsanos2002} models the deviation of
the PSF as uncorrelated zero mean Gaussian noise. The authors of \cite{Molina2006} considered an extension of this
model to a non-sparse, simultaneous autoregression (SAR) prior model for both the image and point spread function.
{ Compared to \cite{Molina2006}, recent papers on single motion deblurring in photography
\cite{Fergus2006,Shan2008} use heavier tailed distributions for the gradient of images and an exponential distribution
for the PSF. In addition, the algorithm in \cite{Fergus2006} separately identifies the PSF using a multi-scale approach
to perform conventional image restoration. The authors of \cite{Shan2008} proposed an image prior to reduce ringing
artifacts from blind deconvolution of photo images. } This paper considers an alternative model, appropriate to MRFM
{but significantly different from photography}, that imposes sparsity on the image and an empirical Bayes prior on
the PSF.

To mitigate the effects of PSF mismatch on MRFM sparse image reconstruction, a non-Bayesian alternating minimization
(AM) algorithm \cite{Herrity2008b} was proposed by Herrity \emph{et al.} which showed robust performance. In this
paper, we propose a hierarchical Bayesian approach to semi-blind image deconvolution that exploits prior information on
the PSF model. This is a semi-blind modification of the Bayesian MRFM reconstruction approach of Dobigeon {\em et al.}
\cite{Dobigeon2009a} that uses an adaptive tuning scheme to produce a Bayesian estimate of the PSF and a Bayesian
reconstruction of the image. The contribution of this paper is a novel Bayesian approach to a joint estimation of PSFs
and images. We represent the PSF on a truncated orthogonal basis, where the basis elements are the singular vectors in
the singular value decomposition of the family of perturbed nominal PSFs. A Gaussian prior model specifies a log
quadratic Bayes prior on deviations from the nominal PSF. Our approach is related to the recent papers of Tzikas
\emph{et al.} \cite{Tzikas2009} and Orieux \emph{et al.} \cite{Orieux2010}. In \cite{Tzikas2009} a pixel-wise,
space-invariant Gaussian kernel basis is assumed with a gradient based image prior. Orieux \emph{et al.} introduced a
Metropolis-within-Gibbs algorithm to estimate the parameters that tune the device response. The strategy
\cite{Orieux2010} focuses on reconstruction with smoothness constraints and requires recomputation of the entire PSF at
each step of the algorithm. This is computationally expensive, especially for complex PSF models such as in the MRFM
instrument. Here, we propose an alternative that consists of estimating the deviation from a given nominal PSF. More
precisely, the nominal point response of the device is assumed known and the true PSF is modeled as a small
perturbation about the nominal response. Since we only need to estimate linear perturbations about the nominal PSF
relative to a low dimensional precomputed and truncated basis set, this leads to reduction in computational complexity
and an improvement in convergence as compared to \cite{Tzikas2009} and \cite{Orieux2010}. We approximate the full
posterior distribution of the PSF and the image using samples generated by a Markov Chain Monte Carlo algorithm.
Simulations and comparisons to other state-of-the-art blind deconvolution algorithms are presented and quantify the
advantages of our algorithm for myopic sparse image reconstruction. We then apply it to real MRFM tobacco virus data
made available by our IBM collaborators.

This paper is organized as follows: Section \ref{sec:problem} formulates the problem. Section \ref{sec:model} covers
the Bayesian framework of image modeling and the following Section \ref{sec:Gibbs} proposes a solution in this
framework. Section \ref{sec:simu} shows simulation results and an application to the real MRFM data.

\section{Forward Imaging and PSF Model} \label{sec:problem}

Let $\MATima$ denote the $l_1\times \ldots \times l_{\dimima}$ unknown $\dimima$-D positive spin density image to be
recovered (e.g., $\dimima=2$ or $\dimima=3$) and $\bfx \in \R^\dimm$ denote the vectorized version of $\bfX$ with
$\dimm =
 l_1 l_2\ldots l_\dimima$.
This image is to be reconstructed from a collection of $\dimn (= \dimm)$ measurements
$\Vobs=\left[\obs{1},\ldots,\obs{\dimn}\right]\transp$ via the following noisy transformation:
\begin{equation}
\label{eq:model_nD}
 \Vobs = \ftrans{\psf}{\bfx} + \Vnoise,
\end{equation}
where $\ftrans{\cdot}{\cdot}$ is the $n$-dimensional convolution operator or the mean response function $\bbE
[\bfy|\psf, \bfx]$, $\Vnoise$ is a $\dimn\times 1$ observation noise vector and $\psf$ is the kernel modeling the
response of the imaging device.

A typical PSF for MRFM is shown in Mamin {\em et al.}\cite{Mamin2003} for horizontal and vertical MRFM tip
configurations. In \eqref{eq:model_nD} $\Vnoise$ is an additive Gaussian noise sequence, independent of $\bfx$,
distributed according to $\Vnoise \sim \calN\left(\Vzero,\noisevar\Id{\dimn}\right)$. The PSF is assumed to be known up
to a perturbation $\Delta\psf$ about a known nominal $\psf_0$:

\begin{equation}
\label{eq:psf}
 \psf = \psf_0 + \Delta\psf.
\end{equation}

In the MRFM application the PSF is described by an approximate parametric function that depends on the experimental
setup. Based on the physical parameters (gathered in the vector $\boldsymbol{\zeta}$) tuned during the experiment
(external magnetic field, mass of the probe, etc.), an approximation $\psf_0$ of the PSF can be derived. However, due
to model mismatch and experimental errors, the true PSF $\psf$ may deviate from the nominal PSF $\psf_0$.

If a vector of the nominal values of parameters $\boldsymbol{\zeta}_0$ for the parametric PSF model
$\psf_{gen}(\boldsymbol{\zeta})$ is known, then direct estimation of a parameter deviation, $\Delta\boldsymbol{\zeta}$,
can be performed by evaluation of $\psf_{gen}(\boldsymbol{\zeta}_0 + \Delta\boldsymbol{\zeta}) $. However, in MRFM, as
shown by Mamin {\em et al.} \cite{Mamin2003}, $\psf_{gen}(\boldsymbol{\zeta})$ is a nonlinear function with many
parameters that are required to satisfy `resonance conditions' to produce a meaningful MRFM PSF. This makes direct
estimation of the PSF difficult.

In this paper, we take a similar approach to the `basis expansions' in \cite[Chap. 5]{StatisticalLearning:2001},
\cite{Tzikas2009} for approximation of the PSF deviation $\Delta\psf$ as linear models. 
this deviation is that $\Delta\psf$ can be expressed as a linear combination of elements of an \emph{a priori} known
basis $\bfv_k$, $k=1,\ldots,K$,
\begin{equation}
  \Delta\psf = \sum_{k=1}^K \psfparam{k} \bfv_k,
\end{equation}
where $\left\{\bfv_k\right\}_{k=1,\ldots,K}$ is a set of basis functions for the PSF perturbations and $\psfparam{k}$,
$k=1,\ldots,K$ are unknown coefficients. To emphasize the influence of these coefficients on the actual PSF, $\psf$
will be denoted $\psf\left(\Vpsfparam\right)$ with $\Vpsfparam= \left[\psfparam{1},\ldots,\psfparam{K}\right]\transp$.
With these notations, \eqref{eq:model_nD} can be rewritten:
\begin{equation}  \label{eq:model}
 \Vobs =\ftrans{\psf(\Vpsfparam)}{\bfx} + \Vnoise = \MATtrans\left(\Vpsfparam\right)\Vima + \Vnoise , \qquad
\end{equation}
where $\MATtrans\left(\Vpsfparam\right)$ is an $\dimn \times \dimm$ matrix that describes convolution by the PSF kernel
$\psf\left(\Vpsfparam\right)$.

We next address the problem of estimating the unobserved image $\Vima$ and the PSF perturbation $\Delta\psf$ under
sparsity constraints given the measurement $\Vobs$ and the bilinear function $\ftrans{\cdot}{\cdot}$.

\section{Hierarchical Bayesian model}
\label{sec:model}
\subsection{Likelihood function}\label{ssec:likelihood}
Under the hypothesis that the noise in \eqref{eq:model_nD} is Gaussian, the observation model likelihood function takes
the form

\begin{equation}
 \label{eq:likelihood}
 f\left(\Vobs | \Vima, \Vpsfparam,\noisevar\right) =
 \left(\frac{1}{2\pi\sigma^2}\right)^{\frac{\dimn}{2}}
 \exp\left(-\frac{\norm{\Vobs-\ftrans{\psf\left(\Vpsfparam\right)}{\Vima}}^2}{2\noisevar}\right),
\end{equation}
where $\norm{\cdot}$ denotes the standard $\ell_2$ norm: $\norm{\bfx}^2=\bfx\transp\bfx$. This likelihood function will
be denoted $f(\bfy| \paramvect)$, where $\paramvect=\left\{\Vima,\Vpsfparam, \noisevar\right\}$.

\subsection{Parameter prior distributions}

In this section, we introduce prior distributions for the parameters $\paramvect$.

\subsubsection{Image prior}
\label{subsubsec:prior_ima} As the prior distribution for $\ima{i}$, we adopt a mixture of a mass at zero and a
single-sided exponential distribution:
\begin{equation}
\label{eq:prior_ima2}
 f\left(\ima{i}|w,a\right) = (1-w) \dirac{\ima{i}} +
\frac{w}{a}\exp\left(-\frac{\ima{i}}{a}\right)\Indicfun{\R_+^*}{\ima{i}},
\end{equation}
where $w \in [0,1], a \in [0,\infty), \delta\left(\cdot\right)$ is the Dirac function,
 ${\R_+^*}$ is a set of real open interval $(0, \infty)$ and
$\Indicfun{\dsE}{x}$ is the indicator function of the set $\dsE$:
\begin{equation}
 \Indicfun{\dsE}{x}=\left\{
           \begin{array}{ll}
            1, & \hbox{if $x\in\dsE$,} \\
            0, & \hbox{otherwise.}
           \end{array}
          \right.
\end{equation}
By assuming the components $\ima{i}$ to be a conditionally independent ($i=1,\ldots,\dimm$) given $w, a$, the following
conditional prior distribution is obtained for the image $\Vima$:
\begin{equation}
\label{eq:prior_Vima}
 f\left(\Vima|w,a\right) = \prod_{i=1}^\dimm\left[(1-w) \delta\left(\ima{i}\right) +
\frac{w}{a}\exp\left(-\frac{\ima{i}}{a}\right)\Indicfun{\R_+^*}{\ima{i}}\right].
\end{equation}
This image prior is similar to the LAZE distribution (weighted average of a Laplacian probability density function
(pdf) and an atom at zero) used, for example, in Ting \emph{et al.} \cite{Ting2006icip,Ting2009}. As motivated by Ting
\emph{et al.} and Dobigeon \emph{et al.} \cite{Ting2009,Dobigeon2009a}, the image prior in (6) has the interesting
property of enforcing some pixel values to be zero, reflecting the natural sparsity of the MRFM images. Furthermore,
the proposed prior in \eqref{eq:prior_ima2} ensures positivity of the pixel values (spin density) to be estimated.

\subsubsection{PSF parameter prior}
We assume that the parameters $\psfparam{1},\ldots,\psfparam{K}$ are \emph{a priori} independent and uniformly
distributed over known intervals associated with error tolerances centered at $0$. Specifically, define the interval
\begin{equation}
 \calS_k
=\left[-\Delta\psfparam{k},\Delta\psfparam{k}\right]
\end{equation}
and define the distribution of $\Vpsfparam$ as

\begin{equation}
 f\left(\Vpsfparam\right) = \prod_{k=1}^{K}\frac{1}{2\Delta\psfparam{k}}
\Indicfun{\calS_k}{\psfparam{k}},
\end{equation}
with $\Vpsfparam=\left[\psfparam{1},\ldots,\psfparam{K}\right]\transp$. In our experiment, $\Delta\psfparam{k}$'s are
set to be large enough to be non-informative, i.e., an improper, flat prior.

\subsubsection{Noise variance prior}
A non-informative Jeffreys' prior is selected as prior distribution for the noise variance:
\begin{equation}
\label{eq:prior_noisevar}
 f\left(\sigma^2\right) \propto \frac{1}{\sigma^2}
\end{equation}
This model is equivalent to an inverse gamma prior with a non-informative Jeffreys' hyperprior, which can be seen by
integrating out the variable corresponding to the hyperprior \cite{Dobigeon2009a}.

\subsection{Hyperparameter priors}
Define the hyperparameter vector associated with the image and noise variance prior distributions as
$\hypervect=\left\{a,w\right\}$. In our hierarchical Bayesian framework, the estimation of these hyperparameters
requires prior distributions in the hyperparameters. These priors are defined in \cite{Dobigeon2009a} but for
completeness their definitions are reproduced below.

\subsubsection{Hyperparameter $a$} A conjugate inverse-Gamma
distribution is assumed for hyperparameter $a$:
\begin{equation}
 a | \Valpha \sim \calI\calG\left(\alpha_0,\alpha_1\right),
\end{equation}
with $\Valpha=\left[\alpha_0,\alpha_1\right]\transp$. The fixed hyperparameters $\alpha_0$ and $\alpha_1$ have been
chosen to produce a vague prior, i.e., $\alpha_0 = \alpha_1 = 10^{-10}$.

\subsubsection{Hyperparameter $w$}
A uniform distribution on the simplex $[0,1]$ is selected as prior distribution for the mean proportion of non-zero
pixels:
\begin{equation}
 w \sim \calU\left([0,1]\right).
\end{equation}

Assuming that the individual hyperparameters are independent the full hyperparameter prior distribution for
$\hypervect$ can be expressed as:
\begin{equation}
\begin{split}
\label{eq:prior_Hyper}
 f\left(\hypervect|\Valpha\right) &=
f\left(w\right)f\left(a\right) \\ &\propto\frac{1}{
 a^{\alpha_0+1}
}\exp\left(-\frac{\alpha_1}{a}\right)\Indicfun{[0,1]}{w} \Indicfun{\mathbb{R}^+}{a} ,
\end{split}
\end{equation}

\subsection{Posterior distribution}

The posterior distribution of $\left\{\paramvect,\hypervect\right\}$ is:
\begin{equation}
\label{eq:fullposterior} f\left(\paramvect,\hypervect|\Vobs\right) \propto
f\left(\Vobs|\paramvect\right)f\left(\paramvect|\hypervect\right)f\left(\hypervect\right),
\end{equation}
with
\begin{equation}
f\left(\paramvect|\hypervect\right) = f\left(\Vima|
 a,w\right)f\left(\Vpsfparam\right)f\left(\sigma^2\right),
\end{equation}
where $f\left(\Vobs|\paramvect\right)$ and $f\left(\hypervect\right)$ have been defined in \eqref{eq:likelihood} and
\eqref{eq:prior_Hyper}. The conjugacy of priors in this hierarchical structure allows one to integrate out the
parameters $\noisevar$, and the hyperparameter $\hypervect$ in the full posterior
distribution~\eqref{eq:fullposterior}, yielding:
\begin{align}
 \label{eq:posterior}
 f & \left(\Vima,\Vpsfparam |\Vobs,\alpha_0,\alpha_1\right) \propto
 \int f\left(\paramvect,\hypervect|\Vobs\right) \mathrm{d}w \space \mathrm{d}a \mathrm{d}\sigma^2 \propto \\
 \nonumber
&  \frac{\calB e\left(1+ n_1, 1 + n_0\right)}{\left\|\Vobs-\ftrans{\psf\left(\Vpsfparam\right)}{\Vima}\right\|^{N}}
\frac{\Gamma\left(n_1 + \alpha_0 \right)}{(\norm{\Vima}_1 + \alpha_1)^{n_1+\alpha_0}}
\prod_{k=1}^{K}\frac{1}{2\Delta\psfparam{k}} \Indicfun{\calS_k}{\psfparam{k}},
\end{align}
where $\calB e$ is the beta function and $\Gamma$ is the gamma function.

The next section presents the Metropolis-within-Gibbs algorithm \cite{Robert1999} that generates samples distributed
according to the posterior distribution $f\left(\Vima,\Vpsfparam|\Vobs\right)$. These samples are then used to estimate
$\Vima$ and $\Vpsfparam$.

\section{Metropolis-within-Gibbs algorithm \\for semi-blind sparse image reconstruction}
\label{sec:Gibbs}

We describe in this section a Metropolis-within-Gibbs sampling strategy that allows one to generate samples
$\left\{\sample{\Vima}{t},\sample{\Vpsfparam}{t}\right\}_{t=1,\ldots}$ distributed according to the posterior
distribution in \eqref{eq:posterior}. As sampling directly from \eqref{eq:posterior} is a difficult task, we will
instead generate samples distributed according to the joint posterior
$f\left(\Vima,\Vpsfparam,\noisevar|\Vobs,\alpha_0,\alpha_1\right)$. Sampling from this posterior distribution is done
by alternatively sampling one of $\Vima,\Vpsfparam,\noisevar$ conditioned on all other variables \cite{Bremaud2001,
Dobigeon2009a}. The contribution of this work to \cite{Dobigeon2009a} is to present an algorithm that simultaneously
estimates both the image and PSF. The algorithm results in consistently stable output images and PSFs.

The main steps of our proposed sampling algorithm are given in subsections~\ref{subsec:sample_Vima}
through~\ref{subsec:sample_noisevar} (see also Algorithm~\ref{algo:Gibbs}).

\begin{algorithm}[h!]
\caption{Metropolis-within-Gibbs sampling algorithm for semi-blind sparse image reconstruction}  \label{algo:Gibbs}
\begin{algorithmic}[1]
  \STATE \emph{\scriptsize{\% Initialization:}}
  \STATE Sample the unknown image $\sample{\Vima}{0}$ from pdf in \eqref{eq:prior_Vima},
  \STATE Sample the noise variance $\samplenoisevar{0}$ from the pdf in \eqref{eq:prior_noisevar},
  \STATE \emph{\scriptsize{\% Iterations:}}
   \FOR{$t=1,2, \ldots, $}
    \STATE Sample hyperparameter $\sample{w}{t}$ from the pdf in
      \eqref{eq:posterior_w},
    \STATE Sample hyperparameter $\sample{a}{t}$ from the pdf in
      \eqref{eq:posterior_a},
    \STATE \label{algostep:sample_ima}For $i=1,\ldots,\dimm$, sample the pixel intensity $\sample{\ima{i}}{t}$ from the
    pdf in
      \eqref{eq:posterior_ima},
    \STATE For $k=1,\ldots,K$, sample the PSF parameter $\sample{\psfparam{k}}{t}$ from the pdf in
      \eqref{eq:posterior_psfparam}, by using Metropolis-Hastings step (see Algo. \ref{algo:Metropolis_Gibbs}),
    \STATE Sample the noise variance $\samplenoisevar{t}$ from the pdf in \eqref{eq:posterior_noisevar},
  \ENDFOR
\end{algorithmic}
\end{algorithm}

\subsection{Generation of samples according to
$f\left(\Vima\left|\Vpsfparam,\noisevar,\Vobs,\alpha_0,\alpha_1\right.\right)$}
\label{subsec:sample_Vima} To generate samples distributed according to
$f\left(\Vima\left|\Vpsfparam,\noisevar,\Vobs,\alpha_0,\alpha_1\right.\right)$, it is convenient to sample according to
$f\left(\Vima, w, a\left|\Vpsfparam,\noisevar,\Vobs,\alpha_0,\alpha_1\right.\right)$ by the following 3-step
procedure.

\subsubsection{Generation of samples according to $f\left(w\left|\Vima,\alpha_0,\alpha_1\right.\right)$}
The conditional posterior distribution of $w$ is
\begin{equation}
 f\left(w\left|\Vima\right.\right) \propto
(1-w)^{n_0+1-1}w^{n_1+1-1},
\end{equation}
where $n_1 = \norm{\Vima}_0$ and $n_0 = \dimm - \norm{\Vima}_0$. Therefore, generation of samples according to
$f\left(w\left|\Vima\right.\right)$ is achieved as follows:
\begin{equation}
\label{eq:posterior_w}
 w|\Vima \sim \calB e\left(1 + n_1, 1 + n_0\right).
\end{equation}

\subsubsection{Generation of samples according to $f\left(a\left|\Vima\right.\right)$}
The joint posterior distribution \eqref{eq:fullposterior} yields:
\begin{equation}
\label{eq:posterior_a}
 a\left|\Vima , \alpha_0, \alpha_1 \right. \sim
\calI\calG\left(\norm{\Vima}_0 + \alpha_0 ,\norm{\Vima}_1 + \alpha_1 \right).
\end{equation}

\subsubsection{Generation of samples according to $f\left(\Vima\left|w,a,\psfparam,\noisevar,\Vobs\right.\right)$}
The posterior distribution of each component $\ima{i}$ ($i=1,\ldots,\dimm$) given all other variables is derived as:
\begin{equation}
  \label{eq:posterior_ima}
 f\left(\ima{i} | w,a,\psfparam,\noisevar,\Vima_{-i},\Vobs\right) \propto
(1-w_i) \delta\left(\ima{i}\right) + w_i \phi_+\left(\ima{i}|\mu_i,\eta^2_i\right),
\end{equation}
where $\Vima_{-i}$ stands for the vector $\Vima$ whose $i$th component has been removed and $\mu_i$ and $\eta^2_i$ are
defined as the following:
\begin{equation}\label{eqApp:mean_var_posterior_ima}
  \eta^2_i = \frac{\noisevar}{\norm{\Vtrans{i}}^2},
  \mu_i = \eta^2_i
      \left(\frac{\Vtrans{i}\transp\bfe_i}{\noisevar}-\frac{1}{a}\right),
\end{equation}
with $\Vtrans{i}$ and $ \bfe_i$ defined in Appendix \ref{appendic:fast_recursive}.

In \eqref{eq:posterior_ima}, $\phi_+\left(\cdot,m,s^2\right)$ stands for the pdf of the truncated Gaussian distribution
defined on $\R^*_+$ with hidden mean $m$ and hidden variance $s^2$. Therefore, from \eqref{eq:posterior_ima}, $\ima{i}
| w,a,\psfparam,\noisevar,\Vima_{-i},\Vobs$ is a Bernoulli truncated-Gaussian variable with parameter $\left(w_i,
\mu_i, \eta^2_i\right)$.

To summarize, generation of samples distributed according to $f\left(\Vima\left|w, \noisevar,a,,\Vobs\right.\right)$
can be performed by updating the coordinates of $\Vima$ using $\dimm$ Gibbs moves (requiring generation of Bernoulli
truncated-Gaussian variables). A pixel-wise fast and recursive sampling strategy is presented in Appendix
\ref{appendic:fast_recursive} and an accelerated sparsity enforcing simulation scheme is described in Appendix
\ref{appendic:selective_sampling}.

\subsection{Generation of samples according to $f\left(\Vpsfparam\left|\Vima,\noisevar,\Vobs\right.\right)$}
 The posterior distribution of the
parameter $\psfparam{k}$ conditioned on the unknown image $\Vima$, the noise variance $\noisevar$ and the other PSF
parameters $\left\{\psfparam{j}\right\}_{j\neq k}$ is
\begin{equation}\label{eq:posterior_psfparam}
 f\left(\psfparam{k}|\Vpsfparam_{-k},\Vima, \noisevar,\Vobs\right) \propto
 \exp\left[-\frac{\norm{\Vobs-\ftrans{\psf\left(\Vpsfparam\right)}{\Vima}}^2}{2\noisevar}\right]
 \Indicfun{\calS_k}{\psfparam{k}},
\end{equation}
with $\Vpsfparam_{-k} = \left\{\psfparam{j}\right\}_{j\neq k}$. We summarize in Algorithm~\ref{algo:Metropolis_Gibbs} a
procedure for generating samples distributed according to the posterior in \eqref{eq:posterior_psfparam} using a
Metropolis-Hastings step with a random walk proposition \cite{Robert1999} from a centered Gaussian distribution. In
order to sample efficiently, the detailed procedure of how to choose an appropriate value of the variance $s^2_k$ of
the proposal distribution for $\psfparam{k}$ is described in Appendix \ref{appendic:acc_rate}. At iteration $t$ of the
algorithm, the acceptance probability of a proposed state $\psfparam{k}^\star$ is:
\begin{equation}
\label{eq:rho}
 \rho_{\psfparam{k}^{(t)} \rightarrow \psfparam{k}^\star} = \min\left(1,
a_k\Indicfun{\calS_k}{\psfparam{k}^\star}\right),
\end{equation}
with
\begin{equation} \label{eq:acceptance_rate}
  2\noisevar \log a_k =
 \norm{\Vobs-\ftrans{\psf\left (\psfparam{k}^{ (t) }\right) }{\Vima}}^2
- \norm{\Vobs-\ftrans{\psf\left (\psfparam{k}^{\star}\right) }{\Vima}}^2.
\end{equation}
Computing the transformation $\ftrans{\cdot}{\cdot}$ at each step of the sampler can be computationally costly.
Appendix \ref{appendic:fast_recursive} provides a recursive strategy to efficiently sample according to
$f\left(\Vpsfparam\left|\Vima,\noisevar,\Vobs\right.\right)$.

\begin{algorithm}[h!]
\caption{Sampling according to $f\left(\psfparam{k}|\Vpsfparam_{-k},\Vima, \noisevar,\Vobs\right)$}
\label{algo:Metropolis_Gibbs}
\begin{algorithmic}[1]
    \STATE Sample $\varepsilon \sim \calN\left(0,s^2_p\right)$,
    \STATE Propose $\psfparam{k}^\star$ according to $\psfparam{k}^\star = \psfparam{k}^{(t)} + \varepsilon$,
    \vspace{0.2cm}
    \STATE Draw $u\sim\calU\left([0,1]\right)$,
    \STATE
      Set $\psfparam{k}^{(t+1)} =
      \left\{\begin{array}{ll}
          \psfparam{k}^\star, & \hbox{if $u \leq \rho_{\psfparam{k}^{(t)} \rightarrow \psfparam{k}^\star}$,} \\
          \psfparam{k}^{(t)}, & \hbox{otherwise.}
        \end{array}
       \right.$
\end{algorithmic}
where $\calU\left(\mathbb{E}\right)$ stands for the uniform distribution on the set $\mathbb{E}$.\vspace{0.2cm}
\end{algorithm}

\subsection{Generation of samples according to $f\left(\noisevar\left|\Vima,\Vobs,\Vpsfparam\right.\right)$}
\label{subsec:sample_noisevar} Samples $(\sigma^2)^{(t)}$ are generated according to the inverse gamma posterior
\begin{equation}
  \label{eq:posterior_noisevar}
 f(\noisevar\left|\Vima,\Vobs,\Vpsfparam\right) =
\calI\calG\left(\frac{\dimn}{2},\frac{\norm{\Vobs - \ftrans{\psf(\Vpsfparam)}{\Vima}}^2}{2}\right).
\end{equation}

\section{Experiments}
\label{sec:simu}

In this section we present simulation results that compare the proposed  semi-blind Bayesian deconvolution algorithms
of Section \ref{sec:Gibbs} with the non-blind method \cite{Dobigeon2009a}, the AM algorithm \cite{Herrity2008b}, and
other blind deconvolution methods. Here a nominal PSF $\psf_0$ was selected that corresponds to the mathematical MRFM
point response model proposed by Mamin \emph{et al.} \cite{Mamin2003}.

Using our MCMC algorithm  described in Sec. \ref{sec:Gibbs}, the MMSE estimators of image and PSF are approximated by ensemble averages over the generated
samples after the burn-in period. The joint MAP estimator is selected among the drawn samples, after the stationary
distribution is achieved, such that it maximizes the posterior distribution $f\left(\Vima,\Vpsfparam |\Vobs\right)$
\cite{Marin2007}.

\begin{table}[h!]
\renewcommand{\arraystretch}{1.4}
\begin{center}
\caption{Parameters used to compute the MRFM PSF.\label{tab:parameters}}
\begin{tabular}{|l|c|c|}
\hline \multicolumn{2}{|c|}{Parameter} & \multirow{2}{5mm}{Value}\\ \cline{1-2} Description & Name & \\ \hline
Amplitude of external magnetic field      & $B_\textrm{ext}$  & $9.4 \times 10^3~\textrm{G}$ \\ Value of
$B_\textrm{mag}$ in the resonant slice & $B_\textrm{res}$  & $1.0 \times 10^4~\textrm{G}$ \\ Radius of tip
& $R_0$        & $4.0~\textrm{nm}$ \\ Distance from tip to sample           & $d_0$         & $6.0~\textrm{nm}$ \\
Cantilever tip moment              & $m$         & $4.6 \times 10^5~\textrm{emu}$\\ Peak cantilever oscillation
oscillation     & $x_\textrm{pk}$   & $0.8~\textrm{nm}$\\ Maximum magnetic field gradient         & $G_\textrm{max}$  &
$125$ \\ \hline
\end{tabular}
\end{center}
\end{table}

\subsection{Simulation on simulated sparse images}
\label{sec:sim1} 
\begin{figure}[h!]
 \centering
 \subfigure[Sparse true image ($\|\bfx\|_0 = 11$)]{\label{fig:Xtrue}
  \includegraphics[width=\figwidthsmall]{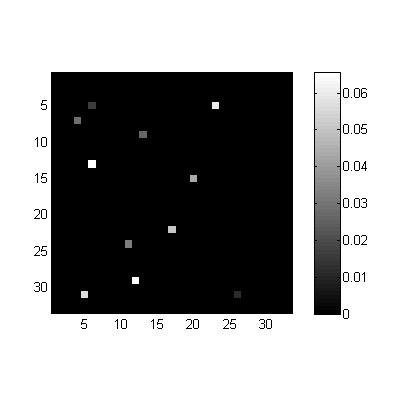}}
 \subfigure[Raw MRFM observation]{\label{fig:obs}
  \includegraphics[width=\figwidthsmall]{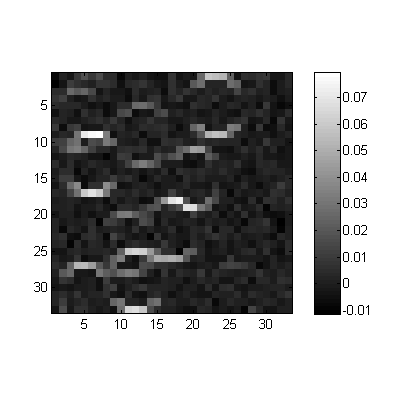}}
 \caption{Simulated true image and MRFM raw image exhibiting superposition of point responses (see Fig.
 \ref{fig:kernel}) and noise.}
 \label{fig:trueobs}
\end{figure}

\begin{figure}[h!]
 \centering
 \subfigure[Nominal MRFM PSF]{\label{fig:Hnominal}
  \includegraphics[width=\figwidthsmall]{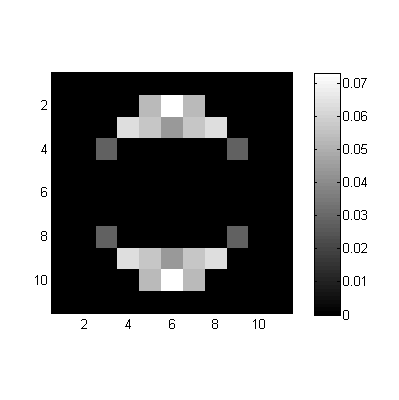}}
 \subfigure[True MRFM PSF]{\label{fig:Htrue}
  \includegraphics[width=\figwidthsmall]{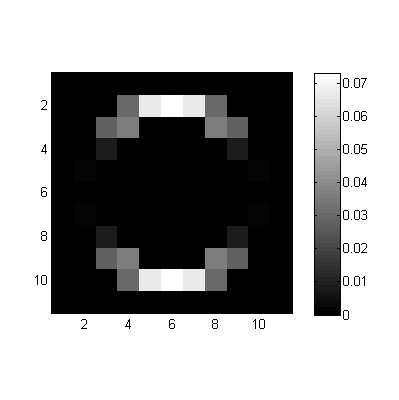}}
 \caption{Assumed PSF $\psf_0$ and actual PSF $\psf$.}
 \label{fig:kernel}
\end{figure}

We performed simulations of MRFM measurements for PSF and image models similar to those described in
Dobigeon \emph{et al.} \cite{Dobigeon2009a}. The signal-to-noise ratio was set to $\textrm{SNR} = 10$dB. Several $32
\times 32$ synthetic sparse images, one of which is depicted in Fig.~\ref{fig:Xtrue}, were used to produce the data and
were estimated using the proposed Bayesian method. The assumed PSF $\psf_0$ is generated following the physical model
described in Mamin \emph{et al.} \cite{Mamin2003} when the physical parameters are tuned to the values displayed in
Table~\ref{tab:parameters}. This yields a $11 \times 11$ $2$-dimensional convolution kernel represented in
Fig.~\ref{fig:Hnominal}. We assume that the true PSF $\psf$ comes from the same physical model where the radius of the
tip and the distance from tip to sample have been mis-specified with $2\%$ error as $R=R_0-2\%=3.92$ and $d = d_0+2\% =
6.12$. This leads to the convolution kernel depicted in Fig.~\ref{fig:Htrue}. The observed measurements $\Vobs$, shown
Fig.~\ref{fig:obs}~ are a $32 \times 32$ image of size $\dimn=1024$.

\begin{figure}[h!]
 \centering
  \includegraphics[width=\figwidthbis]{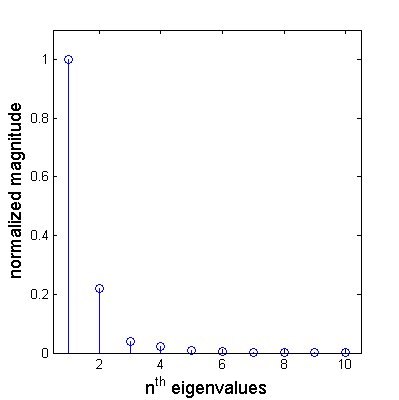}
 \caption{Scree plot of residual PCA approximation error in $l_2$ norm (magnitude is normalized up to the largest
 value, i.e. $\lambda_{max} : = 1$).}
 \label{fig:scree1}
\end{figure}

\begin{figure}[h!]
 \centering
  \includegraphics[width=\figwidth]{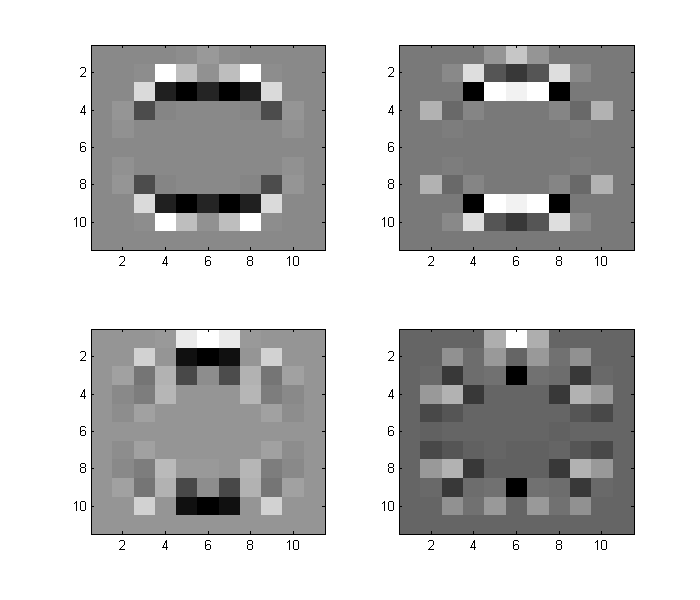}
 \caption{The $K=4$ principal vectors ($\bfv_k$) of the perturbed PSF, identified by PCA.}
 \label{fig:patterns}
\end{figure}

The proposed algorithm requires the definition of $K$ basis vectors $\bfv_k$, $k=1,\ldots,K$, that span a subspace
representing possible perturbations $\Delta\psf$. We empirically determined this basis using the following PSF
variational eigendecomposition approach. A set of $5000$ experimental PSFs $\tilde\psf_j$, $j=1,\ldots,5000$, were
generated following the model described in Mamin \emph{et al.} \cite{Mamin2003} with parameters $d$ and $R$ randomly
drawn according to Gaussian distribution\footnote{ We used a PSF generator provided by Dan Rugar's group at IBM
\cite{Mamin2003}. The variances of the Gaussian distributions are carefully tuned so that their standard deviations
produce a minimal volume ellipsoid that contains the set of valid PSF's of the form specified in \cite{Mamin2003}. }
centered at the nominal values $d_0$, $R_0$, respectively. Then a standard principal component analysis (PCA) of the
residuals $\left\{\tilde\psf_j-\psf_0\right\}_{j=1,\ldots,5000}$ , by allowing the maximum variance over the parameters
that produce valid MRFM PSFs, was used to identify $K=4$ principal axes that are associated with the basis vectors
$\bfv_k$. The necessary number of basis vectors, $K=4$ here, was determined empirically by detecting a knee at the
scree plot shown in Fig.~\ref{fig:scree1}. The first four eigenfunctions, corresponding to the first four largest
eigenvalues, explain $98.69\%$ of the observed perturbations. The 4 principal patterns of basis vectors are depicted in
Fig.~\ref{fig:patterns}.

The proposed semi-blind Bayesian reconstruction algorithm was applied to estimate both the sparse image and the PSF
coefficients of $\bfv_k$'s, using the prior in \eqref{eq:prior_ima2}. From the observation shown in Fig. \ref{fig:obs}
the PSF estimated by the proposed algorithm is shown in Fig.~\ref{fig:Hest} and is in good agreement with the true one.
The corresponding maximum \emph{a posteriori} estimate (MAP) of the unknown image is depicted in
Fig.~\ref{fig:Xmyopic}. The obtained coefficients of the PSF-eigenfunctions are close to true coefficients (Fig.
\ref{fig:HestCurve}).

\subsection{Comparison to Other Methods}
\label{ssec:comparison}

\begin{figure}[h!]
 \centering
 \subfigure[MAP, proposed method ]{\label{fig:Xmyopic}
  \includegraphics[width=\figwidthsmall]{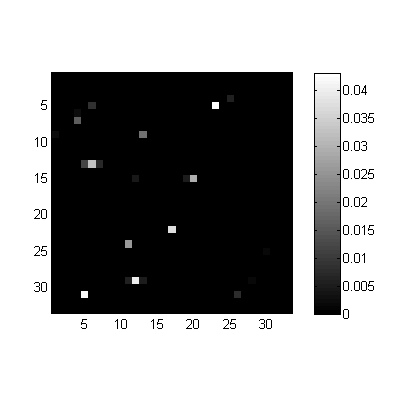}}
 \subfigure[MAP, Bayesian non-blind method with $\psf_0$]{\label{fig:XpreMC}
  \includegraphics[width=\figwidthsmall]{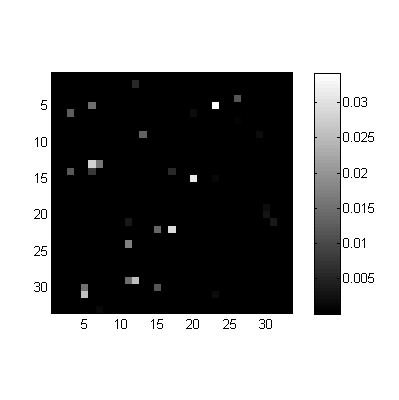}}
 \subfigure[AM ]{\label{fig:XpreAM}
  \includegraphics[width=\figwidthsmall]{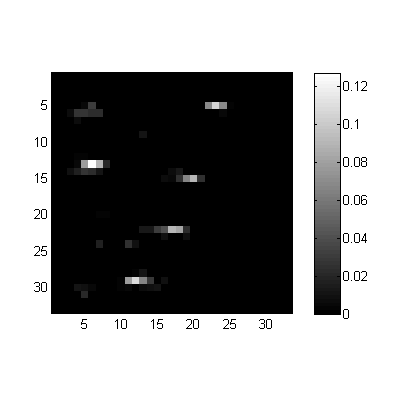}}
 \subfigure[Almeida's method]{\label{fig:X_almeidas}
  \includegraphics[width=\figwidthsmall]{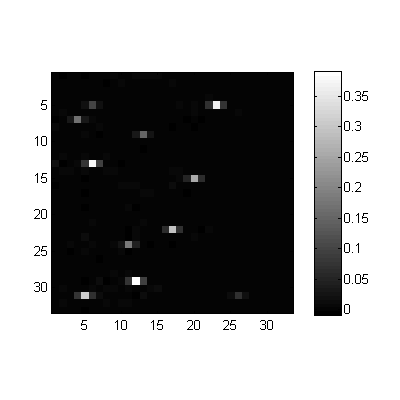}}
 \subfigure[Tzikas' method ]{\label{fig:X_tzikas}
  \includegraphics[width=\figwidthsmall]{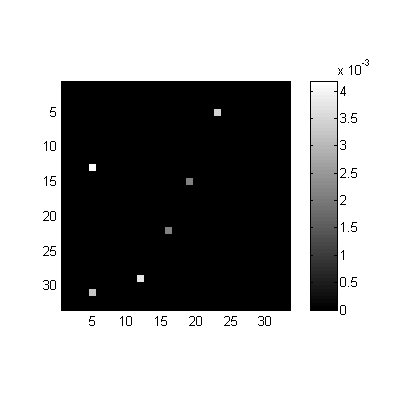}}
 \caption{The true sparse image and estimated images from Bayesian non-blind, AM, our semi-blind, Almeida's, and
 Tzikas' methods.}
 \label{fig:Xest}
\end{figure}

\begin{figure}[h!]
  \centering
 \includegraphics[width=\figwidth]{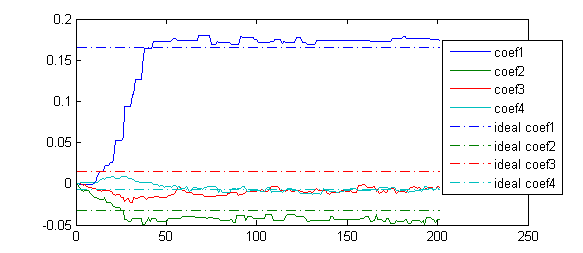}
 \caption{Estimated PSF coefficients for 4 PCs over 200 iterations. These curves show fast convergence of our
 algorithm. `Ideal coefficients' are the projection values of the true PSF onto the space spanned by four principal PSF
 basis.}
 \label{fig:HestCurve}
\end{figure}

For comparison to a non-blind method, Fig.~\ref{fig:XpreMC} shows the estimate using
 the Bayesian non-blind method \cite{Dobigeon2009a} with a mismatched PSF.
Fig.~\ref{fig:XpreAM} shows the estimate generated by the AM algorithm. The nominal PSF described in Section
\ref{sec:sim1} is used in the AM algorithm and hereafter in other semi-blind algorithms, and the parameter values of AM
algorithm were set empirically according to the procedure in \cite{Herrity2008b}. Our proposed algorithm appears to
outperform the others (Fig.~\ref{fig:Xest}) while preserving fast convergence (Fig.~\ref{fig:HestCurve}).

\begin{figure}[h!]
 \centering
 \subfigure[Proposed method]{\label{fig:Hest}
  \includegraphics[width=\figwidthsmall]{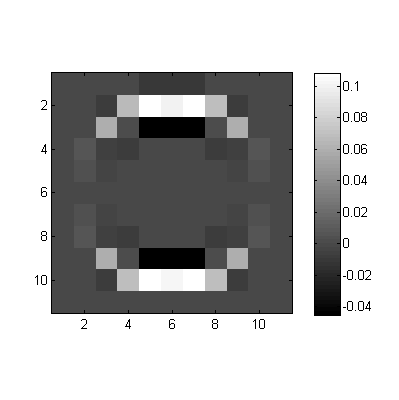}}
  \subfigure[Amizic's method]{\label{fig:PSF_amizic}
  \includegraphics[width=\figwidthsmall]{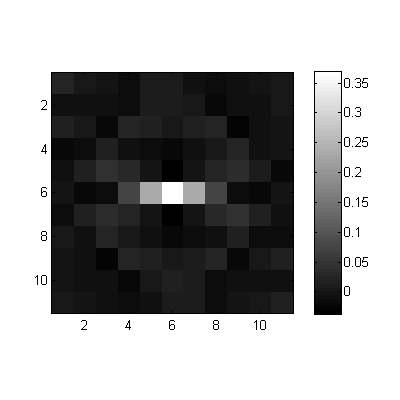}}
 \subfigure[Almeida's method]{\label{fig:PSF_almeidas}
  \includegraphics[width=\figwidthsmall]{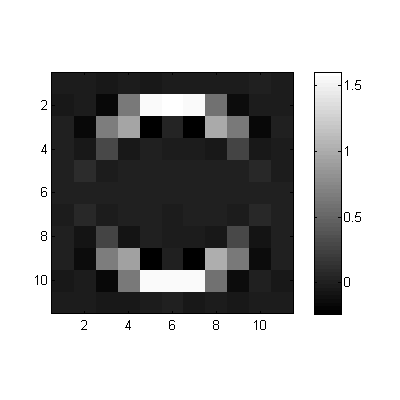}}
 \subfigure[Tzikas' method]{\label{fig:PSF_tzikas}
  \includegraphics[width=\figwidthsmall]{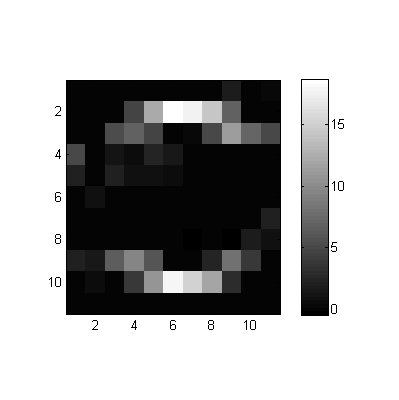}}
 \caption{Estimated PSF $\hat \psf$ of MRFM tip using our semi-blind method, Amizic's method (using TV prior),
 Almeida's method (using progressive regularization), and Tzikas' method (using the kernel basis PSF model),
 respectively. For fairness, we used sparse image priors for the methods. (See Section \ref{ssec:comparison} for
 details on the methods.)}
 \label{fig:kernel_est}
\end{figure}

\begin{figure}[h!]
 \centering
 \subfigure[Histograms of the normalized $l_2$ norm errors. $x$-axis is $\| \frac{\bfx}{\|\bfx\|} - \frac{\hat
 \bfx}{\|\hat \bfx\|} \|_2^2/\|\bfx\|_0$.]{\label{fig:hist_err2}
  \includegraphics[width=\figwidthhalf]{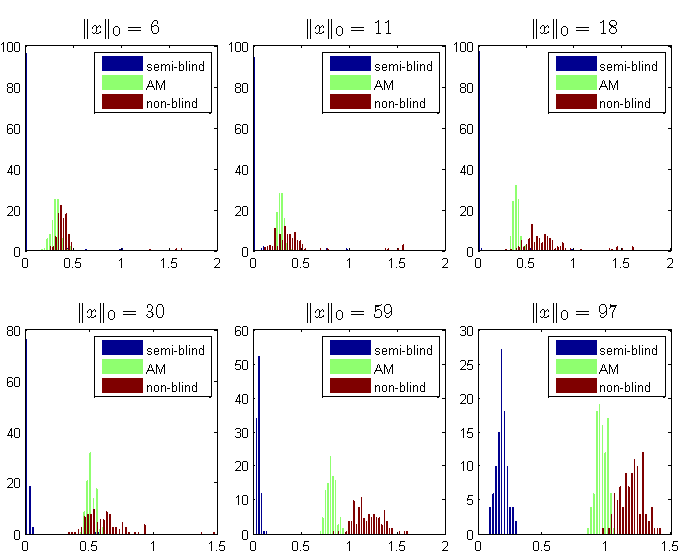}}
 \subfigure[Histograms of the $l_0$ measures. $x$-axis is $\|\hat\bfx\|_0$.
 ]{\label{fig:hist_norm0}
  \includegraphics[width=\figwidthhalf]{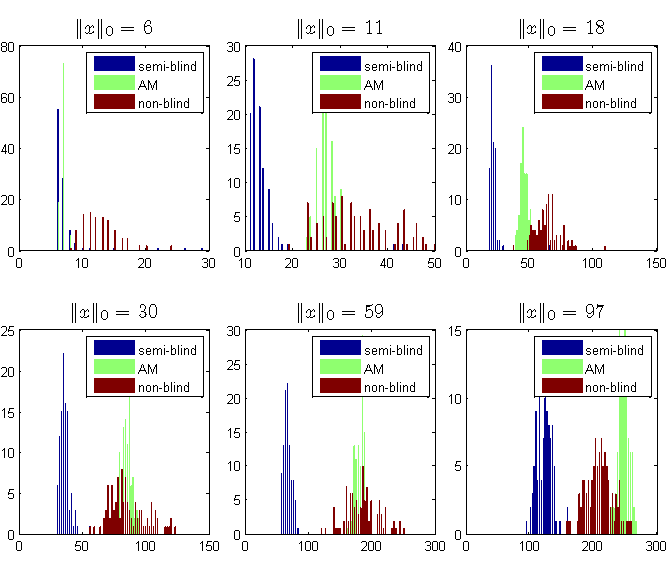}}
 \caption{Histograms of $l_2$ and $l_0$ norm of the reconstruction error. Note that the proposed semi-blind
 reconstructions exhibit smaller mean error and more concentrated error distribution than the non-blind method of
 \cite{Dobigeon2009a} and the alternating minimization method of \cite{Herrity2008b}. }
 \label{fig:histogram}
\end{figure}

Quantitative comparisons were obtained by generating different noises in $100$ independent trials for a fixed true
image. Here, six different true images with six corresponding different sparsity levels ($\|\bfx\|_0 = 6, 11, 18, 30,
59, 97$) were tested. Fig.~\ref{fig:histogram} presents the two histograms of the results with the six sets in the
corresponding two error criteria, $\| \hat\bfx - \bfx \|^2$, $\| \hat\bfx \|_0$, respectively, both of which indicate
that our method performs better and is more stable than the other two methods.

\begin{figure}[h!]
 \centering
 \subfigure[$\| \hat \bfx \|_0/\|\bfx\|_0$: estimated sparsity. Normalized true level is $1$. ]{\label{fig:norm0Plot}
  \includegraphics[width=\figwidthbig]{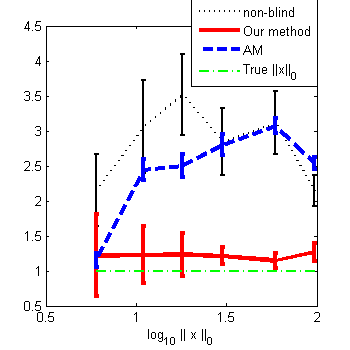}}
 \subfigure[$\| \frac{\bfx}{\|\bfx\|} - \frac{\hat \bfx}{\|\hat \bfx\|} \|_2^2/\|\bfx\|_0$: normalized error in
 reconstructed image]{\label{fig:errtPlot}
  \includegraphics[width=\figwidthbig]{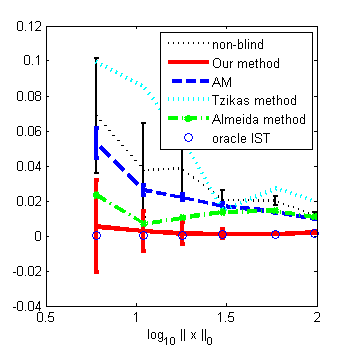}}
 \subfigure[$\| \bfy - \hat \bfy \|_2^2 /\|\bfx\|_0$: residual (projection) error]{\label{fig:errPlot}
  \includegraphics[width=\figwidthbig]{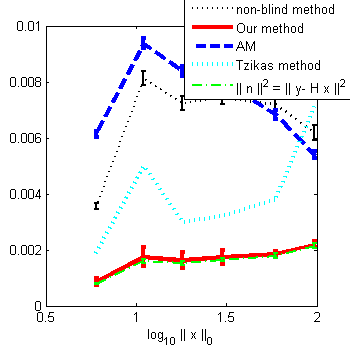}}
 \subfigure
 [$\| \frac{\hat \psf}{\|\hat \psf\|} - \frac{\psf}{\|\psf\|} \|^2_2$, as a performance gauge of our myopic method. At
 the initial stage of the algorithm, $\| \frac{\psf_0}{\|\psf_0\|} - \frac{\psf}{\|\psf\|} \|^2_2 = 0.5627$
 ]{\label{fig:psferrPlot}
  \includegraphics[width=\figwidthbig]{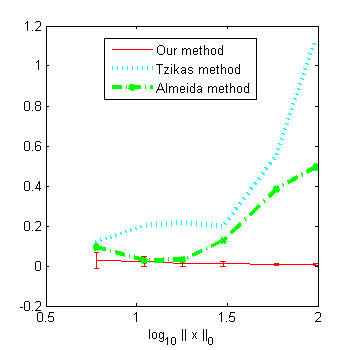}}
 \caption{Error bar graphs of results from our myopic deconvolution algorithm. For several image $\bfx$'s of different
 sparsity levels, errors are illustrated with standard deviations. (Some of the sparsity measure and residual errors
 are too large to be plotted together with results from other algorithms.)}
 \label{fig:MC_est_all}
\end{figure}

Fig.~\ref{fig:MC_est_all} shows reconstruction error performance for several measures of error used in Ting \emph{et
al.} \cite{ Ting2009} and Dobigeon \emph{et al.} \cite{Dobigeon2009a} to compare different reconstruction algorithms
for sparse MRFM images. Notably, compared to the AM algorithm that aims to compensate `blindness' of the unknown PSF
and the previous non-blind method, our method reveals a significant performance gain under most of the investigated
performance criteria and sparsity conditions.


{ In addition to the AM and non-blind comparisons shown in Fig.~\ref{fig:histogram}, we made direct comparisons
between our sparse MRFM reconstruction method and several state-of-the-art blind image reconstruction methods
\cite{Amizic2010,Almeida2010,Tzikas2009, Fergus2006,Shan2008}. In all cases the algorithms were
 initialized with the nominal, mismatched PSF and were applied
 to a sparse MRFM-type image like in Fig.~\ref{fig:trueobs}.
For a fair comparison, we made a sparse prior modification in the image model of other algorithms. The total variation
(TV) based prior for the PSF suggested by Amizic \emph{et al.} \cite{Amizic2010} was also implemented. The obtained PSF
from this method was considerably worse than the one estimated by our proposed method (see Fig.~\ref{fig:PSF_amizic})
resulting in a very poor quality image deconvolution\footnote{ Because this PSF is wrongly estimated and similar to the
2$D$ delta function, the image estimate looks similar to the denoised version of observation, so we omit the image
estimate.}.}

{ The recent blind deconvolution method proposed by Almeida \emph{et al.} \cite{Almeida2010} utilizes the `sharp
edge' property in natural images, with initial, high regularization in order to effectively evaluate the PSF. This
iterative approach has a sequentially decreasing regularization parameter to reconstruct fine details of the image.
Adapted to sparse images, this method performs worse than our method, in terms of image and PSF estimation errors. The
PSF and image estimates from Almeida's method are presented in Fig.~\ref{fig:PSF_almeidas} and \ref{fig:X_almeidas},
respectively. }

{ Tzikas \emph{et al.} \cite{Tzikas2009} uses a similar PSF model to our method using basis kernels. However, no
sparse image prior was assumed in \cite{Tzikas2009} making it unsuitable for sparse reconstruction problems such as the
MRFM problem considered in the paper. For a fair comparison, we applied the suggested PSF model \cite{Tzikas2009} along
with our sparse image prior. The results of PSF and image estimation and the performance graph are shown in
 Fig.~\ref{fig:PSF_tzikas}, Fig.~\ref{fig:X_tzikas}, and Fig. \ref{fig:MC_est_all}, respectively.
In terms of PSF estimation error, our algorithm outperforms the others. }

{ We also compared against the mixture model-based algorithm of Fergus \emph{et al.} \cite{Fergus2006}, and the
related method of Shan \emph{et al.} \cite{Shan2008}, which are proposed for blind deconvolution of shaking/motion
blurs and do not incorporate any sparsity penalization. When applied to the sparse MRFM reconstruction problem the
algorithms of \cite{Fergus2006} and \cite{Shan2008} performed very poorly (produced divergent or trivial solutions; not
shown due to space limitations). This poor performance is likely due to the fact that
 the shape of the MRFM PSF and the sparse image model are significantly different from those in blind deconvolution of
 camera shaking/motion blurs.
The generalized PSF model of \cite{Fergus2006} and \cite{Shan2008} with the sparse image prior is Tzikas' model
\cite{Tzikas2009}, which is described above. }

We used the Iterative Shrinkage/Thresholding (IST) algorithm with a true PSF to lower bound our myopic reconstruction
algorithm. The IST algorithm effectively reconstructs images with a sparsity constraint \cite{Daubechies2004}. From
Fig.~\ref{fig:errtPlot} the performance of our result is as good as that of the oracle IST. In Table \ref{tab:time} we
present comparisons of computation time\footnote{Matlab is used under Windows 7 Enterprise and HP-Z200 (Quad 2.66 GHz)
platform.} of the proposed sparse semi-blind Bayes reconstruction to that of several other algorithms.


\begin{table}[h!]
\caption{\label{tab:time} Computation time of algorithms (in seconds), for the data in Fig.~\ref{fig:trueobs}.
}
\begin{center}
\renewcommand{\arraystretch}{1.2}
\begin{tabular}{|c|c|}
\hline \hline Proposed method   & 19.06 \\ \hline Bayesian nonblind \cite{Dobigeon2009a} & 3.61  \\ \hline IST
\cite{Daubechies2004}  & 0.09\\ \hline
 AM  \cite{Herrity2008b} &  0.40\\
 \hline
  Almeida's method \cite{Almeida2010} &5.63 \\
  \hline
  Amizic's method \cite{Amizic2010} & 5.69\\
  \hline
  Tzikas' method \cite{Tzikas2009} &20.31\\
\hline \hline
\end{tabular}
\end{center}
\end{table}

\subsection{Application to 3D MRFM image reconstruction}
\label{sec:virus} 
\begin{figure}[h!]
  \centering
 \includegraphics[width=\figwidth]{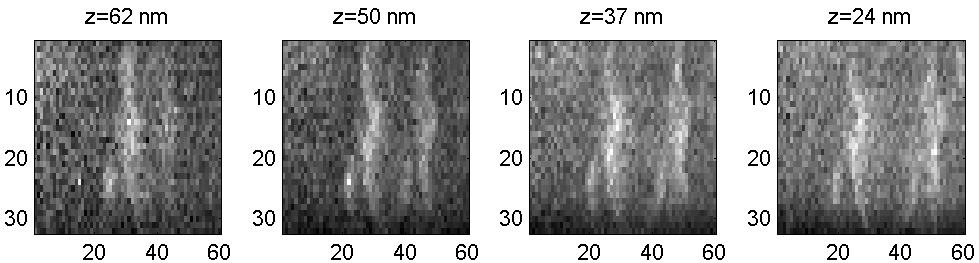}
 \caption{Observed data at various tip-sample distances $z$. }
 \label{fig:virus_obs}
\end{figure}

In this section, we apply the semi-blind Bayesian reconstruction algorithm to the $3$D MRFM tobacco
virus data
 \cite{Degen2009} shown in Fig.~\ref{fig:virus_obs}.
The necessary modification for our algorithm to apply to $3$D data is simple because the extension of our $2$D
pixel-wise sampling method requires only one more added dimension to extend to 3D basis vectors and 3D convolution
kernel. As seen in Appendix \ref{appendic:fast_recursive}, the voxel-wise update of a vectorized image $\bfx$ can be
generalized to $n$D data. This scalability is another benefit of our algorithm. The implementation of AM algorithm is
impractical due to its slow convergence rates \cite{Herrity2008b}. Here we only consider Bayesian methods. The additive
noise is assumed Gaussian consistently with \cite{Rugar1992,Degen2009}, so the noise model in paragraph
\ref{ssec:likelihood} is applied here.

The PSF basis vectors were obtained from a standard PCA and the number of principal components (PC) in the PSF
perturbation was selected as 4 based on detecting the knee in a scree plot. The same interpolation method as used in
\cite{Dobigeon2009a} was adopted to account for unequal spatial sampling rates in the supplied data for the PSF domain
and the image domain.

In the PSF and image domains, along $z$ axis, the grid in PSF signal space is 3 times finer than the observation
sampling density, because the PSF sampling rate along z-axis is 3 times higher than the data sampling rate is. To
interpolate this lower sampled data, we implemented a version of the Bayes MC reconstruction that compensates for
unequal projection sampling in $z$ directions using the interpolation procedure of Dobigeon \emph{et al.}
\cite{Dobigeon2009a}.

\begin{figure}[h!]
 \centering
 \subfigure[Ground truth synthetic virus image obtained from data in Degen {\it et al} \cite{Degen2009}.]{\label{fig:virus_true_synthetic}
  \includegraphics[width=\figwidthsmall]{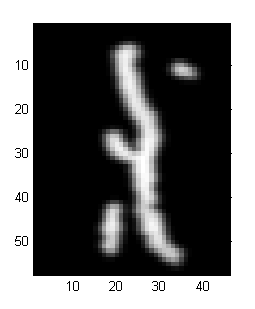}}
 \subfigure[Semi-blind reconstruction of the synthetic virus data. Only the z-planes that have non-zero image intensity are
 shown.)]{\label{fig:virus_synthetic}
  \includegraphics[width=\figwidthsmall]{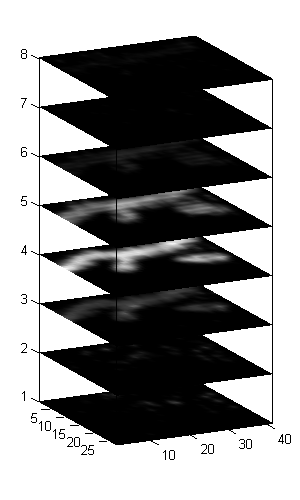}}
 \caption{Results of applying the proposed semi-blind sparse image reconstruction algorithm to synthetic 3D MRFM virus image.}
 \label{fig:virus_synthetics00}
\end{figure}

\begin{figure}[h!]
 \centering
 \subfigure[True PSF. ]{\label{fig:PSF_true_synthetic}
  \includegraphics[width=\figwidthsmall]{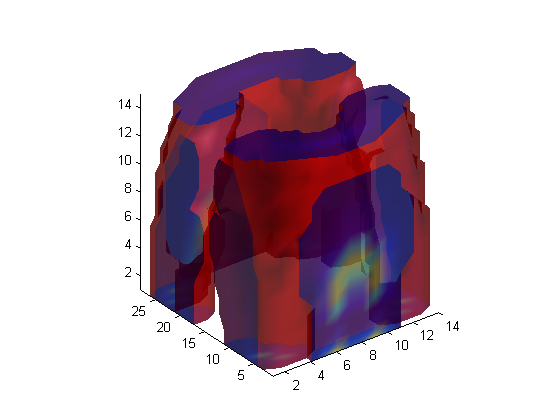}}
 \subfigure[Initial, mismatched PSF.]{\label{fig:PSF_init_synthetic}
  \includegraphics[width=\figwidthsmall]{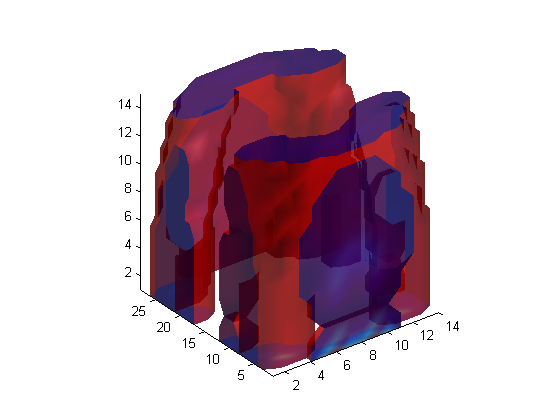}}
 \subfigure[Estimated PSF.]{\label{fig:PSF_est_synthetic}
  \includegraphics[width=\figwidthsmall]{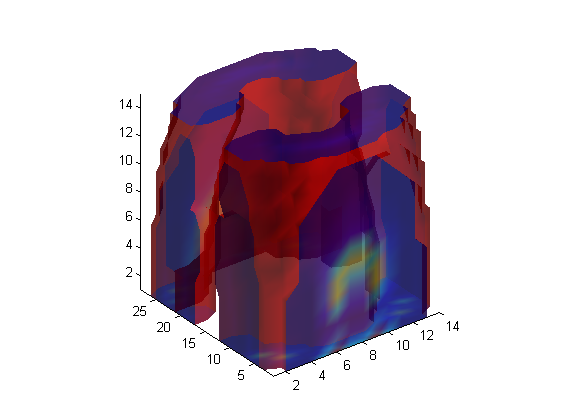}}
 \caption{PSF estimation result.}
 \label{fig:virus_PSF_synthetic00}
\end{figure}

{To demonstrate that the proposed 3D MCMC semi-blind reconstruction algorithm is capable of reconstruction in the presence of significant MRFM PSF mismatch, we first applied it to a simulated version of the experimental data shown in Fig.~\ref{fig:virus_obs}.  We used the scanning electron microscope (SEM) virus image reported in Degen \emph{et al.} \cite{Degen2009} to create a synthetic  3D MRFM virus image, one slice of which is shown in Fig.~\ref{fig:virus_true_synthetic}.
This 3D image was then passed through the MRFM forward model, shown in Fig.~\ref{fig:PSF_true_synthetic}, and 10dB Gaussian noise was added. The mismatched PSF (Fig.~\ref{fig:PSF_init_synthetic}) was used to initialize our proposed semi-blind reconstruction algorithm. After 40 iterations the algorithm reduced the initial normalized PSF error $\| \frac{ \psf_0 } { \|\psf_0\|}
- \frac{ \psf} { \|\psf\|} \|^2$ from $0.7611$ to $\| \frac{ \hat\psf } { \|\hat\psf\|} - \frac{ \psf} { \|\psf\|} \|^2= 0.0295$. Fig.~\ref{fig:virus_synthetic} and Fig.~\ref{fig:PSF_est_synthetic} show the estimated image and the estimated PSF, respectively.}

\begin{figure}[h!]
 \centering
 \subfigure[MAP estimate in $3$D and the estimated image on $6$th plane, showing a virus
 particle.]{\label{fig:virus_myopic01_slice}
  \includegraphics[width=\figwidthhalf]{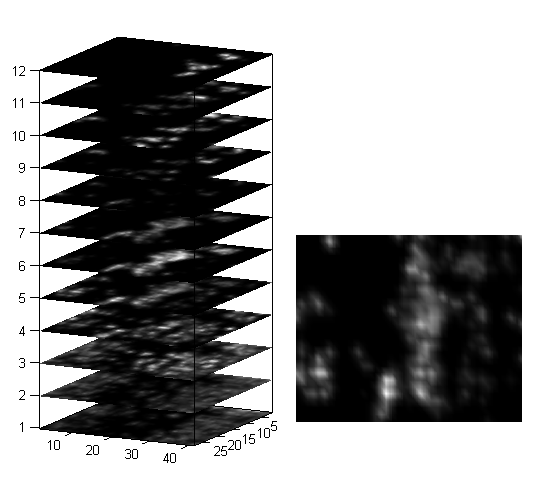}}
 \subfigure[Estimated (left) and nominal (right) PSFs. $ \| \frac{\hat\psf}{\|\hat\psf\|} - \frac{\psf_0}{\| \psf_0 \|}
 \|^2 = 0.0212$. The difference between two is small. (Hard thresholding with level $=\max(PSF) \times 10^{-4}$ is
 applied for visualization. )]{\label{fig:virus_PSF_est}
  \includegraphics[width=\figwidthhalf]{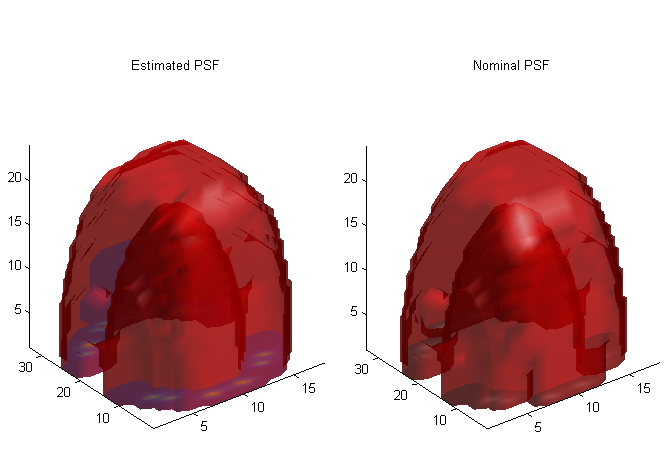}}
 \caption{Semi-blind MC Bayes method results and PSF coefficient curves. $\Delta z = 4.3 nm$, pixel spacing is $8.3nm
 \times 16.6nm$ in $x \times y$, respectively. The size of $(x,y)$ plane is $498nm \times 531.2nm$. Smoothing is
 applied for visualization.}
 \label{fig:virus_est_all}
\end{figure}

\begin{figure}[h!]
 \centering
 \subfigure[MMSE solution. Gray level image intensity values range from 0 (black) to $7.34 \times 10^{-12}$ (white).]{\label{fig:virus_MMSEE}
  \includegraphics[width=\figwidthsmall]{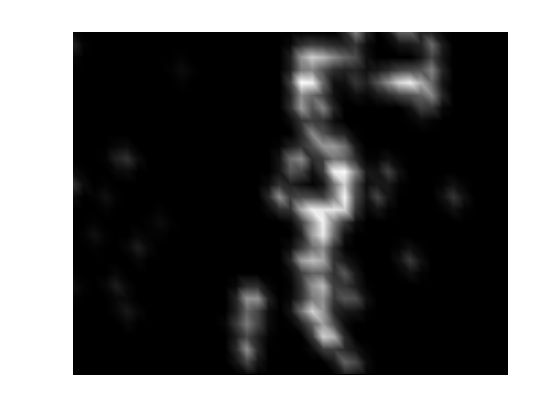}}
 \subfigure[The pixel-wise square root of the image variance. White color indicates a high variance. Gray level image intensity values range from 0 (black) to $3.29 \times 10^{-12}$ (white).]{\label{fig:virus_variance}
  \includegraphics[width=\figwidthsmall]{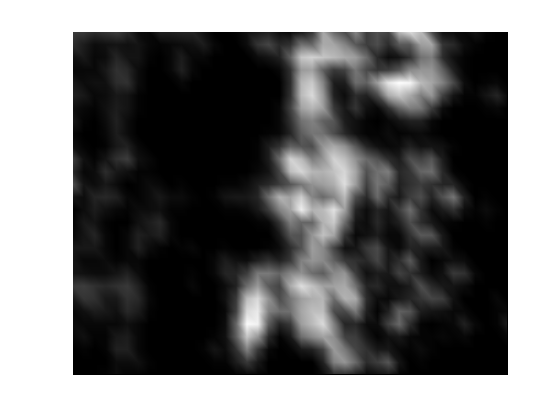}}
 \caption{The posterior mean and variance at the 6th plane of the estimated image (Fig.~ \ref{fig:virus_myopic01_slice}). }
 \label{fig:virus_MMSEE_var}
\end{figure}

{
We next applied the proposed semi-blind reconstruction algorithm to the actual experimental data shown in Fig.~\ref{fig:virus_obs} for which there is neither ground truth on the MRFM image or on the MRFM PSF.  The image reconstruction results are shown in Fig.~\ref{fig:virus_est_all}. The small difference between the nominal PSF and
the estimated PSF indicates that the estimated PSF is close to the assumed PSF. We empirically validated this small difference by verifying that multiple runs of the Gibbs sampler gave low variance PSF residual coefficients.  We conclude from this finding that the PSF model of Degen \emph{et al.} \cite{Degen2009} is in fact nearly Bayes optimal for this experimental data.  The blind image reconstruction shown in Fig.~\ref{fig:virus_est_all} is similar to the image reconstruction in  Degen \emph{et al.} \cite{Degen2009} obtained from applying the Landweber reconstruction algorithm with the nominal PSF.
\\
Using the MCMC generated posterior distribution obtained from the experimental MRFM data, we generated confidence intervals, posterior mean and posterior variance of the pixel intensities of the unknown virus image. The posterior mean and variance are presented in Fig.~\ref{fig:virus_MMSEE_var} for selected pixels. In addition, to demonstrate the match between the estimated region occupied by the virus particle and the actual region we evaluated Bayesian p-values for object regions. The Bayesian p-value for a specific region $R_i$ having non-zero intensity is
$pv(R_i)=\bbP(\{I_k=1\}_{k\in R_i}|\bfy)$ where $\bbP$ is a probability measure and $I_k$ is an indicator function at
the $k$th voxel. Assuming voxel-wise independence, the p-values are easily computed from the posterior distribution and provide a level of a posteriori confidence in the statistical significance of the reconstruction. We found that over 95\% of the Bayesian p-values were greater than 0.5 for the non-zero regions of the reconstruction.
}

\subsection{Discussion} 
\label{ssec:unicity} Joint identifiability is a common issue underlying all blind deconvolution methods. (e.g., scale
ambiguity.) Even though the unicity of our solution is not proven, given the conditions that 1) $\text{span}(\psf) =
\psf_{0} + \textrm{span}(\sum \psf_i)$ does not cover a kernel of a delta function, $\psf=\delta(\cdot)$, and 2) the
PSF solution is restricted to this linear space of $\psf_0, \psf_i$'s, the equation \eqref{eq:posterior} for the MAP
criteria promises a reasonable solution that is not trivial such as $\hat{\bfx} = \bfy$. Due to this restriction and
the sparse nature of the image to be estimated, we can reasonably expect that the solution provided by the algorithm is
close to the true PSF. A study of unicity of the solution would be worthwhile but is beyond the scope of this paper as
it would require study of the complicated and implicit fixed points of the proposed Bayes objective function.

{ Note that proposed sparse image reconstruction algorithm can be extended to exploit sparsity in other domains,
such as the wavelet domain. In this case, if we define $\bfW$ to be the transformation matrix on $\bfx$, the proposed
semi-blind approach can be applied to reconstruct the transformed signal $\bfz = \bfW \bfx$. However, instead of
assigning the single sided exponential distribution as prior for $\bfz$, a double sided Laplacian distribution might be
used to cover the negative values of the pixels. The estimation procedure for PSF coefficients, noise level, and
hyperparameters would be the same. For image estimation, the vector $\bfh_i$ used in
\eqref{eqApp:mean_var_posterior_ima} would be replaced with the $i$th column of $\bfH \bfW^{-1}$. }

\section{Conclusion}
\label{sec:conclusions}

We have proposed an extension of the method of the non-blind Bayes reconstruction in Dobigeon \emph{et al.}
\cite{Dobigeon2009a} that simultaneously estimates a partially known PSF and the unknown but sparse image. The method
uses a prior model on the PSF that reflects a nominal PSF and uncertainty about the nominal PSF. In our algorithm the
values of the parameters of the convolution kernel were estimated by a Metropolis-within-Gibbs algorithm, with an
adaptive mechanism for tuning random-walk step size for fast convergence. Our approach can be used to empirically
evaluate the accuracy of assumed nominal PSF models in the presence of model uncertainty. In our sparse reconstruction
simulations, we demonstrated that the semi-blind Bayesian algorithm has improved performance as compared to the AM
reconstruction and other blind deconvolution algorithms and non-blind Bayes method under several criteria.

Possible extensions of the proposed method may include enforcing sparsity constraints on the result PSF and the
eigenfunctions, by using sparse PCA type algorithms. Also, even with the selective sampling strategy that speeds up the
sampling, the MCMC methods are slower than nonstochastic methods. This will be addressed in future work.

\section*{Acknowledgement}
The authors gratefully acknowledge Dr. Dan Rugar for providing the tobacco virus data and his insightful comments on
this work. { They also thank the reviewers for the comments, which helped the authors significantly improve the
quality of this paper. }

\appendices

\section{Fast recursive sampling strategy}
\label{appendic:fast_recursive} In iterative MRFM algorithms such as AM and the proposed Bayesian method, repeated
evaluations of the transformation $\ftrans{ \psf\left(\Vpsfparam\right) }{\Vima}$ can be computationally difficult. For
example, at each iteration of the proposed Bayesian myopic deconvolution algorithm, one must generate $x_i$ from its
conditional distribution $ f\left(\ima{i} | w,a,\psfparam,\noisevar,\Vima_{-i},\Vobs\right)$, which requires the
calculation of $\ftrans{\psf}{\tilde{\Vima}_i}$ where $\tilde{\Vima}_i$ is the vector $\Vima$ whose $i$th element has
been replaced by $0$. Moreover, sampling according to the conditional posterior distributions of $\noisevar$ and
$\psfparam{k}$ \eqref{eq:posterior_noisevar} and \eqref{eq:posterior_psfparam} requires computations of
$\ftrans{\boldsymbol{\psf}}{\Vima}$.

By exploiting the bilinearity of the transformation $\ftrans{\cdot}{\cdot}$, we can reduce the complexity of the
algorithm. We describe below a strategy, similar to those presented in \cite[App. B]{Dobigeon2009a}, which only
requires computation of $\ftrans{\cdot}{\cdot}$ at most $\dimm\times(K+1)$ times. First, let $\bfI_{\dimm}$ denote the
$\dimm\times\dimm$ identity matrix and $\bfu_i$ its $i$th column. In a first step of the analysis, the $\dimm$ vectors
$\bfh_i^{(0)}$ ($i=1,\ldots,\dimm$)
\begin{equation}
 \bfh_i^{(0)} = \ftrans{\psf_0}{\bfu_i},\
\end{equation}
and $K\dimm$ vectors $\bfh_i^{(k)}$ ($i=1,\ldots,\dimm,\
 k=1,\ldots,K$)
\begin{equation}
 \bfh_i^{(k)} = \ftrans{\bfv_k}{\bfu_i},\
\end{equation}
are computed. Then one can compute the quantity $\ftrans{\psf}{\tilde{\Vima}_i}$ and $\ftrans{\psf}{\Vima}$ at any
stage of the Gibbs sampler without evaluating $\ftrans{\cdot}{\cdot}$, based on the following decomposition
\begin{equation}
 \ftrans{\boldsymbol{\psf}}{\Vima} = \sum_{i=1}^{\dimm} \ima{i}
 \bfh_i^{(0)} + \sum_{k=1}^{K}\psfparam{k}
 \sum_{i=1}^{\dimm}\ima{i}\bfh_i^{(k)}.
\end{equation}

The resulting procedure to update the $i$th coordinate of the vector ${\Vima}$ is described in
Algorithm~\ref{algo:efficient_sim} below. 

\begin{algorithm}[h!]
\caption{Efficient simulation according to $f\left(\Vima \left| w,a,\noisevar,\Vobs\right.\right)$}
\label{algo:efficient_sim} At iteration $t$ of the Gibbs sampler, for $i=1,\ldots,\dimm,$ update the $i$th coordinate
of the vector $$
 \samplebis{\Vima}{t,i-1} =
\left[\ima{1}^{(t)},\ldots,\ima{i-1}^{(t)},\ima{i}^{(t-1)},\ima{i+1}^{(t-1)},\ldots,\ima{\dimm}^{(t-1)}\right]\transp$$
via the following steps:
\begin{algorithmic}[1]
  \STATE compute $\bfh_i =\bfh_i^{(0)}+\sum_{k=1}^{K} \psfparam{k} \bfh_i^{(k)}$,
  \STATE set $ \ftrans{\psf}{\tilde{\Vima}^{(t,i-1)}_i} = \ftrans{\psf}{{\Vima}^{(t,i-1)}} -
        \ima{i}^{(t-1)}\bfh_i$,
  \STATE set $\bfe_i = \Vima -\ftrans{\psf}{\tilde{\Vima}^{(t,i-1)}_i}$,
  \STATE compute $\mu_i$, $\eta^2_i$ and $w_i$ as defined in
      [6],
  \STATE draw $\ima{i}^{(t)}$ according to \eqref{eq:posterior_ima},
  \STATE set $\samplebis{\Vima}{t,i} =
        \left[\ima{1}^{(t)},\ldots,\ima{i-1}^{(t)},\ima{i}^{(t)},\ima{i+1}^{(t-1)},\ldots,\ima{\dimm}^{(t-1)}\right]\transp$,
  \STATE set $\ftrans{\psf}{{\Vima}^{(t,i)}} = \ftrans{\psf}{\tilde{\Vima}^{(t,i-1)}_i} + \ima{i}^{(t)}\bfh_i$.
  \end{algorithmic}
\end{algorithm}

Note that in step 7. of the algorithm above, $\ftrans{\psf}{\Vima}$ is recursively computed. Once all the coordinates
have been updated, the current $\ftrans{\psf}{\Vima}$ can be directly used to sample according to the posterior
distribution of the noise variance in \eqref{eq:posterior_noisevar}. Moreover, this quantity can be used to sample
according to the conditional posterior distribution of $\psfparam{k}$ in \eqref{eq:posterior_psfparam}. More precisely,
evaluating $\ftrans{\psf\left(\psfparam{k}^{\star}\right)}{\Vima}$ in the acceptance probability
\eqref{eq:acceptance_rate} can be recursively evaluated as follows
\begin{equation}
 \ftrans{\psf\left(\psfparam{k}^{\star}\right)}{\Vima} =
 \ftrans{\psf\left(\psfparam{k}^{(t)}\right)}{\Vima} -
 \left(\psfparam{k}^{(t)}-\psfparam{k}^{\star}\right) \sum_{i=1}^{\dimm}
 \ima{i}\bfh_i^{(k)}.
\end{equation}


\section{Sparsity enforcing selective sampling}
\label{appendic:selective_sampling} Since we have estimated the `overall sparsity', $1-\hat w$, of $\bfx$ from
\eqref{eq:posterior_w}, we can expedite the sampling procedure of $\bfx$ by selectively sampling only significant
portions of entire pixels of $\bfx$. As a result, we expect $(1-\hat w) \times 100\%$ of pixel domain of $\bfx$ to be
zero, which will not need to be re-sampled.

At time $t$, in order to approximate the quantile $(1-\hat w)$ of $\{w_i^{(t)}\}_{i=1, \ldots, M}$ in
\eqref{eq:posterior_ima} we evaluate the $(1-\hat w)$ quantile value of the previously obtained $\{w_i^{(t-1)}\}_{i=1,
\ldots, M} $. This approximation accelerates the computation because the exact calculation of $\{w_i^{(t)}\}_{i=1,
\ldots, M}$ requires sampling of all $x_i$'s. Let $q = quantile( \{w_i^{(t-1)}\}_{i=1, \ldots, M}, 1-\hat w)$ and
$w_{thr} = \max (q , 1-\hat w)$. When $w_i^{(t)}$ for $x_i^{(t)}$ from \eqref{eq:posterior_ima} is less than $w_{thr}$,
then $x_i^{(t)}$ is not updated or is set to zero. Because MCMC sampling is computationally expensive, especially for
large size images, this suggestion can be restricted to the burn-in period to save computations.

In our experiment, the selective sampling of $\bfx$ applied after 3 or 4th iterations produce equally good results
compared to the conventional MCMC sampling methods, while reducing computation time by $30 - 50\%$ for non-blind sparse
reconstruction with a fixed PSF and by $10 - 30\%$ for the semi-blind sparse reconstruction.

\section{Adaptive tuning of an acceptance rate in the random-walk sampling}
\label{appendic:acc_rate}

For an efficient sampling of $\lambda_k , k =1,\ldots, K,$ from the desired distribution $\pi(\lambda_k) =
f\left(\psfparam{k}|\Vpsfparam_{-k},\Vima, \noisevar,\Vobs\right)$, we need to properly tune the acceptance rate of the
samples from the proposal distribution. A careful selection of a step size is critical for convergence of the method.
For example, if the step size is too large, most of the iterations will be rejected and the sampling algorithm will be
inefficient. On the other hand, if the step size is too small, most of the random walk moves are accepted but these
moves are slow to cover the probable space of the desired distribution, and the method is once again inefficient.

The transition density of Metropolis-Hastings sampling is $q(\lambda^{(t)},\lambda^{\star (t)})
acc(\lambda^{(t)},\lambda^{\star (t)})$, where $q(\lambda^{(t)},\lambda^{\star (t)})$ is the proposal density from
$\lambda^{(t)}$ and $acc(\lambda^{(t)},\lambda^{\star (t)})$ is the acceptance probability for the move from
$\lambda^{(t)}$ to $\lambda^{\star (t)}$. Here we denote $\lambda_k$ by $\lambda$ without a subscript for simplicity.
We set $q(\lambda^{(t)},\lambda^{\star (t)})$ to be a Gaussian density function of $\lambda^{\star (t)}$, denoted by
$q(\lambda^{(t)},\lambda^{\star (t)}) = q(\lambda^{\star (t)}-\lambda^{(t)}) = \phi(\lambda^{\star
(t)};\lambda^{(t)},s^2)$ with a mean $\lambda^{(t)}$ and a variance $s^2$, which produces a random walk sample path.
Since $q(\cdot,\cdot)$ is symmetrical, $acc_s (\lambda^{(t)},\lambda^{\star (t)}) =
\min\left(1,\dfrac{\pi(\lambda^{\star (t)})q(\lambda^{\star
(t)},\lambda^{(t)})}{\pi(\lambda^{(t)})q(\lambda^{(t)},\lambda^{\star (t)}) } \right) = \min\left(1,\dfrac{
\pi(\lambda^{\star (t)})}{\pi(\lambda^{(t)}) } \right) = \rho_{ \lambda^{(t)} \rightarrow \lambda^{\star (t)}}$, as
derived in \eqref{eq:rho}. Then the acceptance probability from a parameter value $\lambda^{(t)}$ is $acc_s
(\lambda^{(t)}) = \int_{\lambda^{\star (t)}} q(\lambda^{(t)},\lambda^{\star (t)}) acc_s (\lambda^{(t)},\lambda^{\star
(t)}) d\lambda^{\star (t)}$. The acceptance rate with a scale parameter $s$, acting as a step size, can be expressed as
$acc_s = \int_{\lambda} \pi(\lambda) acc_s(\lambda) d\lambda$.

We evaluate these integrations by using Monte Carlo methods; $acc_s \approx \frac{1}{n_1}\sum_{t=1}^{n_1}
acc_s(\lambda^{(t)})$ with $\lambda^{(t)} \sim \pi(\lambda^{(t)})$, and $acc_s(\lambda^{(t)}) \approx
\frac{1}{n_2}\sum_{t=1}^{n_2} acc_s(\lambda^{(t)}, \lambda^{\star (t)})$ with $\lambda^{\star (t)} \sim
q(\lambda^{(t)},\lambda^{\star (t)})$. In practice, this empirical version of the integration value is evaluated as
\begin{equation}
\label{eq:acc} acc_s \approx \frac{1}{W}{\sum_{t=1}^W acc_s (\lambda^{(t)} , \lambda^{\star (t)}}),
\end{equation}
after the burn-in period. Therefore we can evaluate the acceptance rate with $s$ by averaging the Boolean variables of
$acc_s (\lambda^{(t)} , \lambda^{\star (t)}), t=1, \ldots, W$, over a time-frame window of length $W$ with realizations
$\{\lambda^{(t)} , \lambda^{\star (t)}\}_t$. In short, we iteratively update $s$ to achieve an appropriate acceptance
rate, $acc_s$, as described in Algorithm \ref{algo:tuning_s}:

\begin{algorithm}[h!]
\caption{Tuning $s$ in the Gaussian proposal density $q(\cdot,\cdot)$} \label{algo:tuning_s} Select upper and lower
limits $acc_H$ and $acc_L$. At each time $t= W, 2W, 3W, \ldots$, tune $s$ via the following steps:
\begin{algorithmic}[1]
    \STATE Evaluate $acc_s$ using \eqref{eq:acc} for the given time-frame window,
    \STATE
      Update $s \leftarrow
      \left\{\begin{array}{ll}
          s \times c, & \hbox{if $acc_s > acc_H$,} \\
          s \div c, & \hbox{if $acc_s < acc_L$,} \\
          s, & \hbox{otherwise.}
        \end{array}
       \right.$
\end{algorithmic}
\end{algorithm}

In practice, we fix the variance of the instrumental distribution at the end of a burn-in period. Consequently, the
transition kernel will be fixed and this guarantees both ergodicity and stationary distribution. In our experiment, we
set a conservative acceptance range, $acc_H = 0.6, acc_L = 0.4$, referring to \cite{Robert1999}, and $W = 20, c=4$.
This strategy can also be applied to the direct parameter estimation described in Appendix
\ref{appendic:PSF_para_sample}.

\section{Direct sampling of PSF parameter values}
\label{appendic:PSF_para_sample}

As described in Section \ref{sec:intro}, in the MRFM experiments, the direct estimation of PSF parameters is difficult
because of the nonlinearity of $\psf_{gen}$ and the slow evaluation of $\psf_{gen}(\boldsymbol{\zeta}')$ given a
candidate value $\boldsymbol{\zeta}'$. However, if $\psf_{gen}$ is simple and evaluated quickly, then a direct sampling
of parameter values can be performed. To apply this sampling method, instead of calculating a linearized convolution
kernel, $\psf\left(\Vpsfparam\right)$, we evaluate the exact model PSF, $\psf_{gen}\left(\boldsymbol{\zeta}\right)$, in
\eqref{eq:posterior_psfparam} and \eqref{eq:acceptance_rate}. Also the proposed parameter vector
$\boldsymbol{\zeta}^\star$ correspondingly replaces a coefficient vector $\lambda^\star$ and the updated PSF is used in
the estimation of other variables. This strategy turns out to be similar to the approach adopted by Orieux \emph{et
al.} in \cite{Orieux2010}.

\bibliographystyle{ieeetran}
\bibliography{biblio}

\begin{thebibliography}{10}
\providecommand{\url}[1]{#1}
\csname url@samestyle\endcsname
\providecommand{\newblock}{\relax}
\providecommand{\bibinfo}[2]{#2}
\providecommand{\BIBentrySTDinterwordspacing}{\spaceskip=0pt\relax}
\providecommand{\BIBentryALTinterwordstretchfactor}{4}
\providecommand{\BIBentryALTinterwordspacing}{\spaceskip=\fontdimen2\font plus
\BIBentryALTinterwordstretchfactor\fontdimen3\font minus
  \fontdimen4\font\relax}
\providecommand{\BIBforeignlanguage}[2]{{%
\expandafter\ifx\csname l@#1\endcsname\relax
\typeout{** WARNING: IEEEtran.bst: No hyphenation pattern has been}%
\typeout{** loaded for the language `#1'. Using the pattern for}%
\typeout{** the default language instead.}%
\else
\language=\csname l@#1\endcsname
\fi
#2}}
\providecommand{\BIBdecl}{\relax}
\BIBdecl

\bibitem{Sidles1991}
J.~A. Sidles, ``Noninductive detection of single-proton magnetic resonance,''
  \emph{Appl. Phys. Lett.}, vol.~58, no.~24, pp. 2854--2856, June 1991.

\bibitem{Sidles1992}
------, ``Folded stern-gerlach experiment as a means for detecting nuclear
  magnetic resonance in individual nuclei,'' \emph{Phys. Rev. Lett.}, vol.~68,
  no.~8, pp. 1124--1127, Feb 1992.

\bibitem{Sidles1995}
J.~A. Sidles, J.~L. Garbini, K.~J. Bruland, D.~Rugar, O.~Z\"uger, S.~Hoen, and
  C.~S. Yannoni, ``Magnetic resonance force microscopy,'' \emph{Rev. Mod.
  Phys.}, vol.~67, no.~1, pp. 249--265, Jan 1995.

\bibitem{Rugar1992}
D.~Rugar, C.~S. Yannoni, and J.~A. Sidles, ``Mechanical detection of magnetic
  resonance,'' \emph{Nature}, vol. 360, no. 6404, pp. 563--566, Dec. 1992.

\bibitem{Zuger1996}
O.~Z\"uger, S.~T. Hoen, C.~S. Yannoni, and D.~Rugar, ``Three-dimensional
  imaging with a nuclear magnetic resonance force microscope,'' \emph{J. Appl.
  Phys.}, vol.~79, no.~4, pp. 1881--1884, Feb. 1996.

\bibitem{Degen2009}
C.~L. Degen, M.~Poggio, H.~J. Mamin, C.~T. Rettner, and D.~Rugar, ``Nanoscale
  magnetic resonance imaging,'' \emph{Proc. Nat. Academy of Science}, vol. 106,
  no.~5, pp. 1313--1317, Feb. 2009.

\bibitem{Chao2004}
S.~Chao, W.~M. Dougherty, J.~L. Garbini, and J.~A. Sidles, ``Nanometer-scale
  magnetic resonance imaging,'' \emph{Review Sci. Instrum.}, vol.~75, no.~5,
  pp. 1175--1181, April 2004.

\bibitem{Zuger1993}
O.~Z\"uger and D.~Rugar, ``First images from a magnetic resonance force
  microscope,'' \emph{Applied Physics Letters}, vol.~63, no.~18, pp.
  2496--2498, 1993.

\bibitem{Zuger1994}
------, ``Magnetic resonance detection and imaging using force microscope
  techniques,'' \emph{J. Appl. Phys.}, vol.~75, no.~10, pp. 6211--6216, May
  1994.

\bibitem{Degen2009compl}
C.~L. Degen, M.~Poggio, H.~J. Mamin, C.~T. Rettner, and D.~Rugar, ``Nanoscale
  magnetic resonance imaging. {S}upporting information,'' \emph{Proc. Nat.
  Academy of Science}, vol. 106, no.~5, Feb. 2009.

\bibitem{Ting2009}
M.~Ting, R.~Raich, and A.~O. Hero, ``Sparse image reconstruction for molecular
  imaging,'' \emph{IEEE Trans. Image Processing}, vol.~18, no.~6, pp.
  1215--1227, June 2009.

\bibitem{Dobigeon2009a}
N.~Dobigeon, A.~O. Hero, and J.-Y. Tourneret, ``Hierarchical {B}ayesian sparse
  image reconstruction with application to {MRFM},'' \emph{IEEE Trans. Image
  Processing}, vol.~18, no.~9, pp. 2059--2070, Sept. 2009.

\bibitem{Mamin2003}
J.~Mamin, R.~Budakian, and D.~Rugar, ``Point response function of an {MRFM}
  tip,'' IBM Research Division, Tech. Rep., Oct. 2003.

\bibitem{Makni2004}
S.~Makni, P.~Ciuciu, J.~Idier, and J.-B. Poline, ``Semi-blind deconvolution of
  neural impulse response in {fMRI} using a {G}ibbs sampling method,'' in
  \emph{Proc. IEEE Int. Conf. Acoust., Speech, and Signal (ICASSP)}, vol.~5,
  May 2004, pp. 601--604.

\bibitem{Pillonetto2007}
G.~Pillonetto and C.~Cobelli, ``Identifiability of the stochastic semi-blind
  deconvolution problem for a class of time-invariant linear systems,''
  \emph{Automatica}, vol.~43, no.~4, pp. 647--654, April 2007.

\bibitem{Sarri1998}
P.~Sarri, G.~Thomas, E.~Sekko, and P.~Neveux, ``Myopic deconvolution combining
  {K}alman filter and tracking control,'' in \emph{Proc. IEEE Int. Conf.
  Acoust., Speech, and Signal (ICASSP)}, vol.~3, 1998, pp. 1833--1836.

\bibitem{Chenegros2007}
G.~Chenegros, L.~M. Mugnier, F.~Lacombe, and M.~Glanc, ``{3D} phase diversity:
  a myopic deconvolution method for short-exposure images: application to
  retinal imaging,'' \emph{J. Opt. Soc. Am. A}, vol.~24, no.~5, pp.
  1349--1357, May 2007.

\bibitem{Molina1994}
R.~Molina, ``On the hierarchical {B}ayesian approach to image restoration:
  applications to astronomical images,'' \emph{IEEE Trans. Pattern Analysis and
  Machine Intelligence}, vol.~16, no.~11, pp. 1122 --1128, nov 1994.

\bibitem{Galatsanos2000}
N.~P. Galatsanos, V.~Z. Mesarovic, R.~Molina, and A.~K. Katsaggelos,
  ``Hierarchical {B}ayesian image restoration from partially known blurs,''
  \emph{IEEE Trans. Image Processing}, vol.~9, no.~10, pp. 1784 --1797, oct
  2000.

\bibitem{Galatsanos2002}
N.~P. Galatsanos, V.~Z. Mesarovic, R.~Molina, A.~K. Katsaggelos, and J.~Mateos,
  ``Hyperparameter estimation in image restoration problems with partially
  known blurs,'' \emph{Optical Eng}, vol.~41, pp. 1845--1854, 2002.

\bibitem{Molina2006}
R.~Molina, J.~Mateos, and A.~K. Katsaggelos, ``Blind deconvolution using a
  variational approach to parameter, image, and blur estimation,'' \emph{IEEE
  Trans. Image Processing}, vol.~15, pp. 3715--3727, 2006.

\bibitem{Fergus2006}
R.~Fergus, B.~Singh, A.~Hertzmann, S.~T. Roweis, and W.~T. Freeman, ``Removing
  camera shake from a single photograph,'' in \emph{ACM SIGGRAPH 2006 Papers},
  ser. SIGGRAPH '06.\hskip 1em plus 0.5em minus 0.4em\relax New York, NY, USA:
  ACM, 2006, pp. 787--794.

\bibitem{Shan2008}
Q.~Shan, J.~Jia, and A.~Agarwala, ``High-quality motion deblurring from a
  single image,'' \emph{ACM Trans. Graphics}, vol.~27, no.~3, p.~1, 2008.

\bibitem{Herrity2008b}
K.~Herrity, R.~Raich, and A.~O. Hero, ``Blind reconstruction of sparse images
  with unknown point spread function,'' in \emph{Proc. Computational Imaging
  Conference in IS\&T\/SPIE Symposium on Electronic Imaging Science and
  Technology}, C.~A. Bouman, E.~L. Miller, and I.~Pollak, Eds., vol. 6814,
  no.~1.\hskip 1em plus 0.5em minus 0.4em\relax San Jose, CA, USA: SPIE, Jan.
  2008.

\bibitem{Tzikas2009}
D.~Tzikas, A.~Likas, and N.~Galatsanos, ``Variational {B}ayesian sparse
  kernel-based blind image deconvolution with {S}tudent's-t priors,''
  \emph{Trans. Image Processing}, vol.~18, no.~4, pp. 753 --764, April 2009.

\bibitem{Orieux2010}
F.~Orieux, J.-F. Giovannelli, and T.~Rodet, ``Bayesian estimation of
  regularization and point spread function parameters for {W}iener-{H}unt
  deconvolution,'' \emph{J. Opt. Soc. Am. A}, vol.~27, no.~7, pp. 1593--1607,
  July 2010.

\bibitem{StatisticalLearning:2001}
T.~Hastie, R.~Tibshirani, and J.~Friedman, \emph{{The Elements of Statistical
  Learning: Data Mining, Inference, and Prediction}}.\hskip 1em plus 0.5em
  minus 0.4em\relax Springer, Aug. 2003.

\bibitem{Ting2006icip}
M.~Ting, R.~Raich, and A.~O. Hero, ``Sparse image reconstruction using sparse
  priors,'' in \emph{Proc. IEEE Int. Conf. Image Processing (ICIP)}, Oct. 2006,
  pp. 1261--1264.

\bibitem{Robert1999}
C.~P. Robert and G.~Casella, \emph{Monte Carlo Statistical Methods}.\hskip 1em
  plus 0.5em minus 0.4em\relax New York, NY, USA: Springer-Verlag, 1999.

\bibitem{Bremaud2001}
P.~Bremaud, \emph{Markov Chains: Gibbs fields, Monte Carlo Simulation, and
  Queues.}\hskip 1em plus 0.5em minus 0.4em\relax Springer Verlag, 1999.

\bibitem{Marin2007}
J.-M. Marin and C.~P. Robert, \emph{Bayesian Core: A Practical Approach to
  Computational {B}ayesian Statistics}.\hskip 1em plus 0.5em minus 0.4em\relax
  New York, NY, USA: Springer, 2007.

\bibitem{Amizic2010}
B.~Amizic, S.~D. Babacan, R.~Molina, and A.~K. Katsaggelos, ``Sparse {B}ayesian
  blind image deconvolution with parameter estimation,'' in \emph{European
  Signal Processing Conference, Eusipco 2010}.\hskip 1em plus 0.5em minus
  0.4em\relax Aalborg (Denmark), August 2010, pp. 626--630.

\bibitem{Almeida2010}
M.~Almeida and L.~Almeida, ``Blind and semi-blind deblurring of natural
  images,'' \emph{Trans. Image Processing}, vol.~19, no.~1, pp. 36 --52, jan.
  2010.

\bibitem{Daubechies2004}
I.~Daubechies, M.~Defrise, and C.~De~Mol, ``An iterative thresholding algorithm
  for linear inverse problems with a sparsity constraint,''
  \emph{Communications on Pure and Applied Mathematics}, vol.~57, no.~11, pp.
  1413--1457, 2004.

\end{thebibliography}

\end{document}